\documentclass[useAMS,usenatbib,usegraphicx]{mn2e}
\usepackage{amsmath,amssymb,color,latexsym}

\usepackage{graphics}

\title[Equilibrium Configurations of Strongly Magnetized Neutron Stars]{Equilibrium Configurations 
of Strongly Magnetized Neutron Stars with Realistic Equations of State}
\author[Kenta Kiuchi and Kei Kotake]{Kenta Kiuchi $^{1}$\thanks{E-mail:
kiuchi@gravity.phys.waseda.ac.jp}
 and Kei Kotake$^{2,3}$\thanks{E-mail:kkotake@th.nao.ac.jp}\\
$^{1}$Science \& Engineering, Waseda University, 3-4-1 Okubo, Shinjuku,
Tokyo, 169-8555, Japan\\
$^{2}$Division of Theoretical Astronomy, National Astronomical Observatory
  Japan, 2-21-1, Osawa, Mitaka, Tokyo, 181-8588, Japan\\
$^{3}$Max-Planck-Institute f\"{u}r Astrophysik, Karl-Schwarzshild-Str. 1, 
D-85741, Garching, Germany}
\begin{document}

\date{}


\maketitle


\begin{abstract}
We investigate equilibrium sequences of magnetized rotating stars
with four kinds of realistic equations of state (EOSs) of SLy (Douchin
et al.), FPS (Pandharipande et al.), Shen (Shen et
al.), and LS (Lattimer \& Swesty). 
Employing the Tomimura-Eriguchi
scheme to construct the equilibrium configurations.
we study the basic physical properties of the sequences 
in the framework of Newton gravity. In addition we newly take into
account a general relativistic effect to the magnetized rotating 
configurations.
With these computations, we find that the 
properties of the Newtonian magnetized stars, 
e.g., structure of magnetic field, highly depends on the EOSs. 
 The toroidal magnetic fields concentrate rather 
near the surface for Shen and LS EOSs than those for SLy and
FPS EOSs. The poloidal fields are also affected by the
toroidal configurations.
Paying attention to the stiffness of the EOSs, we analyze this
tendency in detail. In the general relativistic stars, we find that the
difference due to the EOSs becomes small because all the employed EOSs
become sufficiently stiff for the large maximum density, typically
greater than $10^{15}\rm{g}~\rm{cm}^{-3}$. The maximum baryon mass of the
magnetized stars with axis
ratio $q\sim 0.7$ increases about 
up to twenty percents for that of spherical stars. We furthermore
compute equilibrium sequences at finite
temperature, which should serve as an initial condition for the 
hydrodynamic study of newly-born magnetars. Our results suggest
that we may obtain information about the EOSs from the observation of
the masses of magnetars.
\end{abstract}


\section{Introduction }\label{intro}
Recently there is a growing evidence of the supermagnetized
neutron stars with the magnetic fields of $\sim 10^{14}-10^{15}$ G, 
the so-called magnetars (e.g., 
\citet{wod,Lattimer:2006}). 
Although such stars are estimated to be 
only a subclass ($\sim 10 \%$) of the canonical neutron stars 
with $\sim 10^{12}-10^{13}$ G
\citep{kouve}, much attention has
been paid to this objects because there remain many
astrophysically exciting but unresolved problems.
Giant flaring activities observed during the past $\sim$20 years 
have given us good opportunities to study the coupling of the
interior to the magnetospheric structures of magnetars 
\citep{Thompson:1995gw,Thompson:1996pe},
but the relationship between the crustal fraction and the subsequent
starquarks has not yet been clarified (see references in \citet{anna}).
The origin of the large magnetic field is also a big problem, whether
generated at post-collapse in the rapidly neutron star
\citep{Thompson:1993hn} or descend from the main sequence stars
\citep{Ferrario:2007bt}.
Assuming the large magnetic fields before core-collapse, 
 extensive magnetohydrodynamic (MHD) stellar collapse
simulations have been done recently
 \citep{ys,kotake,moise,takiwaki,ober,Shibata:2006hr,livne}
in order to understand the formation mechanism of
magnetars with the different levels of the sophistication in the
 treatment of the equations of state (EOSs), the neutrino transport,
 and the general relativity (see \citet{kotake_rev} for a review). 
Here it is worth
mentioning that the gravitational waves from magnetars could be a
useful tool giving us an information about the magnetar interiors 
\citep{Ioka:2001,Bonazzola:1995rb,Kotake:2004ws}.
 While in a microscopic point of view, 
the effects of magnetic field larger than the so-called QED 
limit of $\sim 4.4\times
 10^{13}~\rm{G}$, on the EOSs (e.g., \citet{Lattimer:2006}) and 
the radiation processes 
have been also extensively investigated (see
 \citet{Harding} for a review). 
For the understanding
of the formation and evolution of the magnetars, it is indispensable to
unify these macroscopic and microscopic studies together, albeit 
not an easy job.

When one tries to study the above interesting issues, the
construction of an equilibrium configuration of magnetars may be
one of the most fundamental problems, thus has been extensively
investigated since the pioneering study by  
\cite{ch2}. 
\cite{Ferraro:1954} and \cite{Roberts:1955} studied the equilibrium
configurations of an incompressible 
star with the poloidal fields. 
\cite{Prendergast:1956} then succeeded to take into account the
toroidal fields.
\cite{Woltjer:1960} extended Prendergast's work to the compressible one. 
\cite{Monaghan:1965,Monaghan:1966} and \cite{Roxburgh:1966}  
improved a boundary condition of the poloidal fields. 
Note that all these works mentioned above were based on 
the perturbation approach in which one assumes the magnetic field is 
sufficiently weak. 
\cite{Ostriker:1968} and \cite{Miketinac:1975} developed 
techniques to treat a strong magnetic field beyond the perturbative 
approach with the Newtonian gravity.
\cite{Bocquet:1995}, \cite{Konno:1999zv} and 
\cite{Ioka:2003,Ioka:2004} developed the scheme with the 
general relativistic framework. 

Here it should be noted that the studies  mentioned above 
have contained two drawbacks to be modified. 
In \cite{Prendergast:1956}, \cite{Woltjer:1960}, \cite{Ostriker:1968} 
and \cite{Ioka:2004}, 
the magnetic fields are assumed to vanish at the stellar surface and 
such a condition may be somewhat unrealistic. 
In \cite{Ferraro:1954}, \cite{Roberts:1955}, 
\cite{Monaghan:1965,Monaghan:1966}, \cite{Miketinac:1975}, 
\cite{Bocquet:1995} and \cite{Konno:1999zv}, 
only poloidal fields have been considered. 
It has been pointed out that such a purely poloidal field 
will be unstable and will decay within several Alfv\'{e}n times 
\citep{wri,mark,mark2}. In fact,
\cite{Braithwaite:2004,Braithwaite:2006} recently find by the 3D
numerical simulations that 
the pure poloidal fields certainly decay within several Alfv\'{e}n 
times and the configurations consisted both of poloidal and toroidal ones with 
comparable strength can be only stable.

Recently, \cite{tom} established a new non-perturbative 
method, which has conquered the above drawbacks.
In the scheme, one can obtain a solution with a natural boundary
condition, in which 
the magnetic field vanishes at the infinity. Moreover, one can treat both 
poloidal and toroidal fields simultaneously. 
More recently, \cite{yos} 
calculated a number of sequences of magnetized rotating star and discussed 
their basic properties. 
Extending the Tomimura-Eriguchi scheme to treat the differential
rotation, \cite{yos2} found that the magnetic fields 
can be in a twisted-torus equilibrium structure and discussed the 
universality of such 
structures. 

There may still remain a room to be sophisticated in their studies 
\citep{tom,yos,yos2}
 that the polytropic EOSs have been used for simplicity.  
In general, the central density of the neutron stars is considered to
 be higher than the nuclear density \citep{bp}. Since we still do not have 
an answer about the 
EOS in such a higher regime, many kinds of the nuclear EOSs have been
 proposed (e.g., \cite{Lattimer:2006}) depending
 on the descriptions of the nuclear forces and the possible appearance
 of the exotic physics (e.g., \citet{gled}). 
While the stiffness of the polytropic EOS has been treated to be
globally constant inside the star,
it should depend on the density in the realistic EOSs.
 Remembering that 
the equilibrium configurations are achieved  
by the subtle and local balance of the 
gravitational force, centrifugal force, the Lorentz force, and the pressure 
gradient, 
it is a nontrivial problem how the equilibrium configurations change 
for the realistic EOSs.

Therefore the first purpose in this paper is to extend the studies of
\cite{tom} and \cite{yos} to incorporate the realistic EOSs and
discuss the equilibrium properties.
Four kinds of EOSs of SLy \citep{douchin}, FPS \citep{pand}, Shen
\citep{shen98}, and Lattimer-Swesty \citep{LS} are 
adopted, which are often employed in the recent literatures of the
MHD studies mentioned above. In contrast to the case of the
 polytropic EOS, a maximum density 
enters as a new parameter. At first, we set this value as comparable 
as the nuclear
density of $\rho_{\rm nuc} \approx 2.8\times 10^{14}
\rm{g}~\rm{cm}^{-3}$, because the general relativistic (GR) corrections 
are rather
small at this density regime as will be mentioned.
 If a maximum density of star is much larger than the nuclear density, 
we must take account into a GR effect. 
However, the fully GR approach to the magnetized equilibrium 
configuration has not been established yet except for the purely
 poloidal fields \citep{Bocquet:1995}. Therefore, we consider here
a new approach to take into account a GR effect approximately and 
 see their effects on the equilibrium configurations.
This is the second purpose of this paper. 
Applying this method, 
 we furthermore compute equilibrium sequences at finite
temperature, which should serve as an initial condition for the 
hydrodynamic evolutionary studies of newly-born magnetars. 

This paper is organized as follows. In Section~\ref{basic}, we shall briefly 
summarize the basic equations, the numerical scheme and the employed 
equations of 
state. Numerical results and their analysis are presented 
in Section~\ref{result}. 
Summary and discussion follow in 
Section~\ref{conc}.
\section{Numerical Scheme and Models}\label{basic}
\subsection{Basic equations}
The basic equations describing an equilibrium state of a perfect conductive 
fluid are summarized according to \citet{lov,yos2}:\\
\\
1. Maxwell's equations:
\begin{eqnarray}
\nabla_a E^a = 4 \pi \rho_e,\label{eq1}&&\\
\epsilon^{abe}\nabla_b B_e = \frac{4\pi}{c}j^a,\label{eq2}&&\\
\epsilon^{abe}\nabla_b E_e = 0,\label{eq3}&&\\
\nabla_a B^a = 0,\label{eq4}&&
\end{eqnarray}
where we use Gaussian units, with the charge and current densities 
$\rho_e$ and $j^a$ both measured in electrostatic units. Here $E^a$ , $B^a$, 
and $c$ are the electric field, magnetic field, and the light velocity, 
respectively, $\nabla_a$ stands for the covariant derivative and 
$\epsilon_{abe}$ for the Levi-Civita tensor in flat three-dimensional space.\\
\\
2. Perfect conductivity equations:
\begin{eqnarray}
&& E_a  + \epsilon_{abe} \frac{v^b}{c} B^e = 0,\label{eq5}
\end{eqnarray}
where $v^a$ denotes the fluid velocity.\\
\\
3. Mass conservation:
\begin{eqnarray}
&& \nabla_a(\rho v^a ) = 0,\label{eq6}
\end{eqnarray}
where $\rho$ is the mass density.\\
\\
4. Euler equations:
\begin{eqnarray}
&& \rho v^b \nabla_b v_a + \nabla_a p + \rho \nabla_a \Phi 
-\frac{1}{c} \epsilon_{abe} j^b B^e = 0,\label{eq7}
\end{eqnarray}
where $p$ is the pressure and $\Phi$ is the gravitational potential. Note that 
the last term on the left-hand side of equation (\ref{eq7}) represents the 
Lorentz force being exerted on the perfectly conductive fluid, which can, 
due to equation (\ref{eq2}), be rewritten in terms of $B^a$ as
\begin{eqnarray}
\frac{1}{c}\epsilon_{abe} j^b B^e = \frac{1}{4\pi}( B^b \nabla_b B_a 
- B^b \nabla_a B_b ), \label{eq9}
\end{eqnarray}
5. Poisson equation for the gravitational potential:
\begin{eqnarray}
&& \nabla^a \nabla_a \Phi = 4\pi G \rho,\label{eq10}
\end{eqnarray}
where $G$ is the gravitational constant.

In order to obtain stationary states of magnetized stars, the following 
assumptions are adopted in this frame work:(1) The magnetic axis and the 
rotation axis are aligned. (2) The matter and the magnetic field distributions 
are axisymmetric about this axis. (3) The sources of the magnetic fields, 
i.e., the current distributions are confined to within the stars. For the 
sake of simplicity we require two additional conditions that are not 
necessary to have stationary solutions. (4) The meridional circulation of the 
fluid is neglected. Consequently, the fluid flow can be described by the 
angular velocity $\Omega$ only, which is given by $v^a = \Omega \varphi^a$, 
where $\varphi$ stands for the rotational Killing vector. Due to this 
assumptions and the axisymmetric assumption, the mass conservation 
equation~(\ref{eq6}) 
is automatically satisfied. (5) The barotropic equation of 
state, $p=p(\rho)$, is adopted. Although we employ realistic equations
of state in this paper, this barotropic condition can be maintained as
explained in subsection \ref{EOS}.

Under these assumptions, the basic equations for determining strictures of 
magnetized rotating stars can be derived. For axisymmetric configurations, 
the divergence-free magnetic fields are automatically satisfied by 
introducing $\Psi$ of $R$ and $z$ as follows:
\begin{eqnarray}
&&  B^R = - \frac{1}{R}\frac{\partial}{\partial z} \Psi,\label{eq11}\\
&&  B^z =   \frac{1}{R}\frac{\partial}{\partial R} \Psi.\label{eq12}
\end{eqnarray}
The function $\Psi$ is sometimes called the flux function. Here we have 
employed the cylindrical coordinates $(R,\phi,z)$, in terms of which the 
line element and the Levi-Civita tensor are, respectively, given by
\begin{eqnarray}
&&  ds^2 = g_{ab} dx^a dx^b = dR^2 + R^2 d\phi^2 + dz^2,\label{eq13}\\
&&  \epsilon_{R\phi z} = R.\label{eq14}
\end{eqnarray}
Introducing the vector potential $A_a$, defined as
\begin{eqnarray}
&&  B^a = \epsilon^{abe}\nabla_b A_e,\label{eq15}
\end{eqnarray}
we can confirm that the flux function $\Psi$ is nothing but the $\phi-$ 
component of the vector potential $A_\phi$. Equation (\ref{eq5}) gives us 
the electric fields as
\begin{eqnarray}
&& E_a = - \frac{\Omega}{c} \nabla_a \Psi.\label{eq16}
\end{eqnarray}
Note that $E_\phi = 0$, because $\Psi$ is an axisymmetric function. From 
equations (\ref{eq3}) and (\ref{eq16}) the angular velocity must be expressed 
as 
\begin{eqnarray}
&& \Omega = \Omega(\Psi).\label{eq17}
\end{eqnarray}
This relation is sometimes called Ferraro's law of isorotation~\citep{fer}.\\
\\
The integrability conditions for equation (\ref{eq7}) result in the following 
relations:
\begin{eqnarray}
&&  B^\phi = \frac{S(\Psi)}{R^2},\label{eq18}\\
&&  \frac{j^a}{c} = \frac{\kappa}{4\pi}B^a 
+ \rho(\mu + R^2\Omega\Omega')\varphi^a,\label{eq19}
\end{eqnarray}
where $S$ and $\mu$ are arbitrary functions of $\Psi$, and
\begin{eqnarray}
&& \kappa(\Psi) \equiv S'(\Psi).\label{eq20}
\end{eqnarray}
Here the prime denotes the differentiation with respect to $\Psi$. The 
functional relation $S(\Psi)$ is sometimes called the dynamical torque-free 
condition~\citep{ch,mes}.

From equations (\ref{eq2}) and (\ref{eq19}) we obtain
\begin{eqnarray}
 R\frac{\partial}{\partial R}
\left(\frac{1}{R}\frac{\partial \Psi}{\partial R}\right)
+ \frac{\partial^2 \Psi}{\partial z^2} 
= - 4 \pi \rho R^2 [R^2\Omega\Omega' + \mu(\Psi)]\nonumber\\
  - \kappa(\Psi)S(\Psi),\label{eq21}
\end{eqnarray}
which is generalized Grad-Shafranov equation. The first integral of the 
Euler equation can be expressed as
\begin{eqnarray}
\int \frac{dP}{\rho} = - \Phi + \frac{1}{2}R^2\Omega^2 + \int^\Psi \mu(u)du
 + C,\label{eq22}
\end{eqnarray}
where $C$ is a constant of integration.
Following \cite{tom}, we convert partial differential 
equations (\ref{eq10}) and (\ref{eq21}) into integral equations, given by
\begin{eqnarray}
&&\Phi(r) = - G\int\frac{\rho(\tilde{r})}{|r-\tilde{r}|}d^3\tilde{r},
\label{eq23}\\
&&A_{\phi}(r)\sin \phi = \int\frac{\kappa S + 4\pi (\mu + \tilde{R}^2\Omega
\Omega' )\rho \tilde{R}^2}{4\pi \tilde{R}|r-\tilde{r}|}
\sin\tilde{\phi}d^3\tilde{r},\label{eq24}
\end{eqnarray}
where all boundary conditions for the potentials $\Phi$ and $\Psi$, i.e., 
that the potentials are regular everywhere, are included in the integral 
expressions, and we need not consider them any further. Concerning the 
arbitrary function $\kappa(\Psi)$, since the current vector $j^a$ has to 
vanish outside the star as mentioned before, the function $\kappa$ needs 
to vanish outside the star.

Concerning the functions characterizing the magnetic field distributions, we 
choose the following forms:
\begin{eqnarray}
&& \mu(u) = \mu,\label{eq25}\\
&& S(u)  = \frac{a}{k+1}(u-u_{max})^{k+1}\theta(u-u_{max}),\label{eq26}\\
&& \kappa(u) = S' = a(u-u_{max})^k\theta(u-u_{max}),\label{eq27}
\end{eqnarray}
where $\mu$, $a$, and $k$ are arbitrary constants and $u_{max}$ is the maximum 
value of $\Psi$ in the vacuum region. Here $\theta(x)$ stands for the 
Heaviside's step function, and we have assumed that $k \ge 0 $ in order to 
have 
finite $\kappa$. It should be noted that, as shown below, $\kappa$, defined 
by equation (\ref{eq27}), satisfied the required condition, i.e., $\kappa=0$ 
outside of the star, at least for the stars in the present study.

Since we have not known what rotation law is realized in actual magnetized 
stars, in the present investigation we adopt the rigid rotation,
\begin{eqnarray}
 \Omega(u) = \Omega_0. \label{eq28}
\end{eqnarray}
In order to carry out numerical computations properly, we introduce the 
following non-dimensional variables by using the maximum density $\rho_{max}$ 
and the equatorial radius $r_e$:
\begin{eqnarray}
&& \hat{\rho} \equiv \rho/\rho_{max}\label{eq30},\\
&& \hat{r} \equiv r/r_e = r/\sqrt{\frac{1}{\beta}\frac{p_{max}}
{4\pi G \rho_{max}^2}}\label{eq31},\\
&& \hat{\Omega} \equiv \Omega/\sqrt{4\pi G \rho_{max}}\label{eq32},\\
&& \hat{C} \equiv C / ( 4\pi G \rho_{max} r_e^2 )\label{eq33},\\
&& \hat{\kappa} \equiv \kappa/ \left(\frac{1}{r_e}\right)\label{eq34},\\
&& \hat{\mu} \equiv \mu /\left(\frac{\sqrt{4 \pi G}}{r_e}\right)\label{eq35},\\
&& \hat{A}_{\phi} \equiv A_{\hat{\phi}} / (\sqrt{4 \pi G }\rho_{max} r_e^2)
\label{eq36},\\
&& \hat{B}^a \equiv B_{\hat{a}}/(\sqrt{4 \pi G }\rho_{max} r_e),\label{eq37}
\end{eqnarray}
where $\beta$ is a numerical factor that is introduced to fix the 
non-dimensional equatorial radius to be unity. Here $B_{\hat{a}}$ denotes 
the orthonormal component of $B_a$, which is frequently convenient because 
it has a physical dimension. By using these variables, global physical 
quantities characterizing an equilibrium state, the gravitational potential 
energy $W$, the rotational energy $T$, the internal energy $U$, the magnetic 
energy $H$, the mass $M$ can be expressed as follows:
\begin{eqnarray}
&& \hat{W} = W/( 4\pi G \rho_{max}^2 r_e^5 ),\label{eq38}\\
&& \hat{T} = T/( 4\pi G \rho_{max}^2 r_e^5 ),\label{eq39}\\
&& \hat{U} = U/( 4\pi G \rho_{max}^2 r_e^5 ),\label{eq40}\\
&& \hat{H} = H/( 4\pi G \rho_{max}^2 r_e^5 ),\label{eq41}\\
&& \hat{M} = M/( \rho_{max} r_e^3 ),\label{eq42}
\end{eqnarray}
where
\begin{eqnarray}
&& W \equiv \frac{1}{2}\int_{star} \Phi \rho d^3 r,\label{eq43}\\
&& T \equiv \frac{1}{2}\int_{star} \rho (R\Omega)^2 d^3r,\label{eq44}\\
&& U \equiv \int_{star} p d^3 r,\label{eq45}\\
&& H \equiv - \frac{1}{c}\int_{all~space}x^a \epsilon_{abe} j^b B^e d^3r,
\label{eq46}\\
&& M \equiv \int_{star} \rho d^3r.\label{eq47}
\end{eqnarray}
We employ the Hachisu self-consistent field(HSCF) 
scheme~\citep{hac,tom,yos,yos2}. 
In the HSCF scheme, one of the model 
parameters characterizing an equilibrium star is the axis ratio $q$, defined 
as $q \equiv r_p / r_e$, where $r_e$ is the smallest distance to the surface 
from the origin. With the HSCF scheme, $\rho$, $A_\phi$, $\beta$, $C$, and 
$\Omega_0$ for rotating configurations (for non-rotating configurations) are
iteratively solved, and during iteration cycles, $q$ and other model 
parameters are fixed. 
By changing the axis ratio $q$, and fixing an appropriate set of the 
parameters, we follow one model sequence of equilibrium configurations. In 
actual calculations, we divide the interval $[0,1]$ in the $\hat{r}$
-direction into 100 meshes and the interval $[0,\pi/2]$in the $\theta$-
direction into 200 meshes. Note that it is enough to calculate solutions 
for the interval $[0,\pi/2]$ in the $\theta$-directions because we impose 
the equatorial plane symmetry. The accuracies of the numerical solutions 
are checked with an assessment by the normalized virial equation (e.g.,~
\cite{cow}), defined as
\begin{eqnarray}
VC = | 2T + W + 3U + H |/|W|.\label{virial}
\end{eqnarray}

 For later convenience, here we like to explain about the 
qualitative meanings of the parameters $a$ and $\hat{\mu}$. 
The parameter $\hat{\mu}$ is directly 
involved in the Bernoulli equation~(\ref{eq22}) and plays a role to 
determine the matter distribution. Increasing of 
$a$ enhances magnitude of magnetic field $\Psi$ through the generalized 
Grad-Shafranov equation (\ref{eq21}). Therefore, increasing 
of both $a$ and $\hat{\mu}$ results in an enhancement in the Lorentz force exerting 
on the conductive fluids. Note that $a=0$ means that the magnetic
field has the poloidal component only, however $\hat{\mu}=0$ {\it does not mean } that 
there exists only toroidal component (see equation~(\ref{eq11}), (\ref{eq12}) 
and (\ref{eq18})).

\subsection{Equations of State}\label{EOS}
As mentioned in Section \ref{intro}, equation of state (EOS) is an
important ingredient for determining the equilibrium configurations.
Before conducting an extensive study as done before for the studies of 
rotating equilibrium configurations in which more
variety of EOSs were employed
(e.g., \citet{nozawa,morrison} and references therein), 
we adopt here four kinds of EOSs of 
SLy \citep{douchin}, FPS \citep{pand}, Shen
\citep{shen98}, and Lattimer-Swesty \citep{LS} which are often
employed in the recent MHD studies relevant for magnetars. 

In the study of cold neutron stars, the $\beta$-equilibrium condition
with respect to beta decays of the form $e^{-} + p \longleftrightarrow n +
\nu_{e}$ and $n \longleftrightarrow p + e^{-} + \bar{\nu}_e$,
can be well validated as for the static properties. Since neutrinos
and antineutrinos escape from the star their chemical potentials
vanish at zero-temperature $T=0$ with $T$ being the temperature, 
and the equilibrium condition is $\mu_{\rm n} = \mu_{\rm e} + \mu_{\rm
p}$, with $\mu_{\rm n}$, $\mu_{\rm e}$, and $\mu_{\rm p}$ being the
chemical potentials of neutron, electron, and proton, respectively.
 With the charge neutrality condition, we can determine
the three independent thermodynamic variables, (for example the
pressure as $P(\rho, Y_e, T)$ with
$Y_e$ being the electron fraction), can only be determined by a single
variable, which we take to be the density, namely $P(\rho, Y_e(\rho))$ \citep{shapiro}, noting here that $T=0$
is assumed for the case of the cold neutron stars. Thanks to this, we
can use the formalism mentioned in Section 2 without violating the
barotropic condition of the EOSs. At the maximum densities higher than $\sim 2
\rho_{\rm nuc}$, muons can appear, and higher than $\sim 3
\rho_{\rm nuc}$ \citep{wiringa,akmal}, one may take into possible
appearance of hyperons \citep{gled}.
However, since the muon contribution to pressure at the higher density has
been pointed to be very small \citep{douchin}, and we still do not have detailed
knowledge of the hyperon interactions, we prefer to employ the above
 neutron star matter, namely $e^{-},n, p$, model to higher densities. 
In the following, we shortly summarize features of the
EOSs employed here.

\begin{figure}
  \begin{center}
  \vspace*{40pt}	
    \includegraphics[width=80mm]{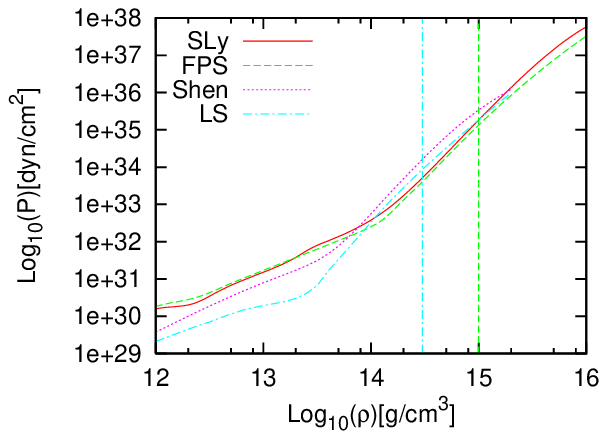}
    \includegraphics[width=80mm]{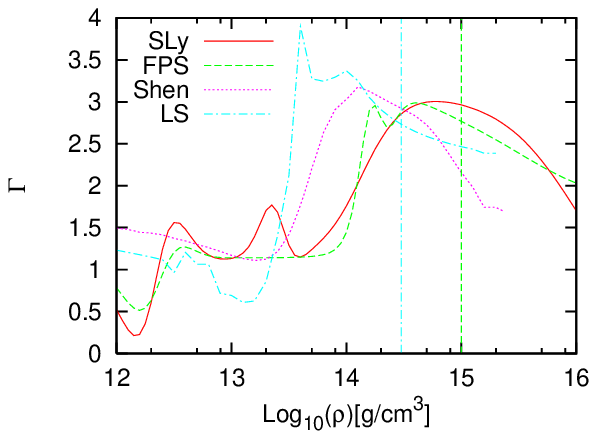}
    \caption{Pressure (top) and the adiabatic index $\Gamma$ (bottom) as a
  function of rest-mass density for SLy \citep{douchin}, FPS \citep{pand}, Shen \citep{shen98}, and LS \citep{LS}
   (LS) EOSs, respectively. The vertical blue (short-dashed) 
   and green (dashed) lines indicate
  the densities of $3\times10^{14}~\rm{g}~\rm{cm}^{-3}$ (Section
  \ref{newton}) and $1\times
10^{15}~\rm{g}~\rm{cm}^{-3}$ (Section \ref{grresult}) set here for the maximum densities. \label{adindex}}
  \end{center}
\end{figure}

Lattimer-Swesty
EOS \citep{LS} is the one which has been used these years as a
standard in the research field of
core-collapse supernova explosions and the subsequent neutron star
formations (see references in \citet{sumi,kotake_rev}), which is based on the compressible drop model for nuclei together
with dripped nucleons. The values of nuclear parameters are chosen
according to nuclear mass formulae and other theoretical studies with
the Skyrme interaction.
Shen EOS \citep{shen98} is the rather modern one currently often used
in the research field, which is based on the relativistic
mean field theory with a local density approximations, which has
 been constructed to reproduce the experimental data of masses and
radii of stable and unstable nuclei (see references in \citet{shen98}).
FPS are modern version of an earlier microscopic EOS calculations 
by \citet{fp81}, which employs both two body (U14) and three-body
interactions (TNI). In the SLy EOS \citep{douchin}, neutron-excess
dependence is added to the FPS EOS, which is more suitable for the
neutron star interiors. As for the FPS and SLy EOSs, we use the fitting formulae 
presented in \cite{Shibata:2005ss}.
In the upper panel of Figure \ref{adindex}, we plot the
pressure $P$ as a function of the rest-mass density $\rho$ for the
EOSs. In the bottom panel, we plot the adiabatic index, defined by
\begin{eqnarray}
\Gamma = \frac{d \ln P}{d \ln \rho}\label{eq:adindex},
\end{eqnarray}
 which can be a useful tool not only to characterize the stiffness but
 also to see their effects on the equilibrium configurations. As seen,
 the adiabatic index as a function of the density is far away from
 constant as in the case of the polytropic EOS, showing an increasing
 (but sometimes zigzag) trend up to the several nuclear density (for example $\sim 2
 \rho_{\rm nuc}$ of SLy) and decreases slowly at higher densities due
 to the interplay of the density dependence of the nuclear
 interactions and of increasing proton fractions. 
Note that the relatively large discrepancies of the pressure below $\sim
10^{14}~\rm{g}~\rm{cm}^{-3}$ and the zigzag features of the adiabatic
 indices are due to the difference in the 
treatment of the inhomogeneous matter consisted of electrons, protons,
free nucleons, and a species of heavy nuclei, and these are pointed
 out to have little effects on determining the equilibrium configurations which
predominantly determined by the behavior of the EOSs at higher
 densities \citep{shenprog}. 

\section{Numerical Results}\label{result}
As mentioned, the first purpose of this study is to investigate the
 basic properties of magnetized rotating stars with 
the four kinds of EOSs in Newtonian gravity. 
In all the calculations, the value of $k$ (equation~(\ref{eq26})) 
is set to be $0.1$ because this choice makes the maximum toroidal
fields to be comparable to the poloidal ones \citep{yos}. 
It should be mentioned that one should specify the maximum density
when using the realistic EOSs, while the degree of freedom of this
choice can be eliminated in case of the polytropic EOSs 
because one can choose the 
polytropic constant freely. We actually fix its value as $3\times
10^{14}~\rm{g}~\rm{cm}^{-3}$ close to the nuclear density.

Before going to the main results, we note that we have checked the accuracy
of our newly developed code by the test problems and their results 
are summarized in appendix \ref{code test}.
\subsection{Newtonian Case}\label{newton}

For clarity, we categorize the equilibrium 
configurations to two types of non-rotating type and 
rotating one. For non-rotating sequence, the parameters, axis ratio $q$ and 
$a$, are input and kept fixed during calculation and then the
parameters, $\beta$, $\hat{C}$, 
and $\hat{\mu}$ are obtained as output. For rotating sequence, $q$, $a$, 
and $\hat{\mu}$ are input and kept fixed, then $\beta$, $\hat{C}$, and 
$\hat{\Omega}_0$ are obtained as outcome. 

\subsubsection{Static Magnetized Configurations \label{SMC}}
We first concentrate on the non-rotating sequences, in which
the anisotropic magnetic stress is only the agent to deform the stars.
In Table~\ref{real-non-rot-mix}, various physical quantities
corresponding to the previous studies \citep{tom,yos,yos2} are given
for the four kinds of EOSs. As a model parameter, we set $a=12$ 
because we are interested in the combination effects of poloidal and
toroidal fields. Note again that the choice of $a=0$ leads to the
 unstable pure poloidal configurations as mentioned. 

As shown in the table, the values of $H/|W|$ for the sequences become large enough
 to be $0.1$ when the stars sufficiently deform typically $q$ of
 the axis ratio to be $ \lesssim 0.7$. In such a region, the perturbative 
approach may break down and a non-perturbative taken here is valid.

\begin{figure}
  \begin{center}
   \includegraphics[width=80mm]{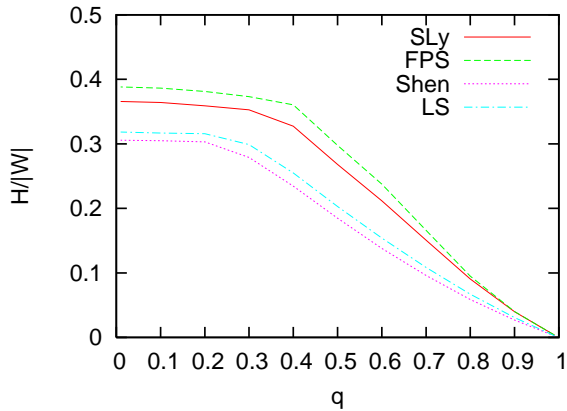}
    \caption{$H/|W|$ as function of $q$ for the non 
rotating sequences shown in Table~
\ref{real-non-rot-mix}.\label{q-H}}
  \end{center}
\end{figure}

Lowering the axis ratio, it can be seen that all sequences can achieve 
nearly toroidal density configuration, $q \sim 0$ due to the strong
Lorentz forces. In fact, $H/|W|$ increases as the
axis ratio $q$ decreases as shown in Figure \ref{q-H}.
This feature does not depend on the EOSs. 
%
Furthermore we find the values of $H/|W|$ for 
SLy or FPS sequences is greater than the those for Shen or LS 
sequences for any axis ratio. 
 This tendency is explained with respect to the stiffness of the 
EOSs. As seen from the bottom panel of Figure 1, Shen or
LS EOSs are 
more stiffer than 
SLy or FPS EOSs in the density region near and below $3\times
10^{14}~\rm{g}~\rm{cm}^{-3}$. 
Therefore, if the stars should have the same axis ratio, the force driven by
the pressure for the Shen or LS stars 
 is greater than the forces for the SLy or FPS stars. 
Consequently, the SLy or FPS stars need much Lorentz force than 
the Shen or LS stars to have the same degree of the deformation.
\begin{table*}
\centering
\begin{minipage}{140mm}
\caption{Physical quantities for the sequence with $a=12$
 and $\hat{\Omega}^2=0$.
\label{real-non-rot-mix}}
\begin{tabular}{ccccccccc}
\hline\hline
q & $H/|W|$ & $U/|W|$ & $|\hat{W}|$ & $\hat{\mu}$ & $\hat{C}$ & 
$\hat{\beta}$ & $ \hat{M} $ & VC \\
\hline
\multicolumn{9}{c}{SLy}\\
\hline
0.99&0.004&0.332&0.140E-01&0.079&-0.311E-01&0.256E-01&0.387&0.125E-03\\
0.90&0.040&0.320&0.168E-01&0.224&-0.374E-01&0.259E-01&0.431&0.130E-03\\
0.80&0.091&0.303&0.212E-01&0.288&-0.468E-01&0.260E-01&0.495&0.112E-03\\
0.70&0.151&0.283&0.277E-01&0.319&-0.597E-01&0.256E-01&0.582&0.828E-04\\
0.60&0.212&0.263&0.376E-01&0.334&-0.772E-01&0.243E-01&0.700&0.690E-04\\
0.50&0.268&0.244&0.460E-01&0.342&-0.935E-01&0.218E-01&0.795&0.813E-04\\
0.40&0.328&0.224&0.377E-01&0.342&-0.905E-01&0.184E-01&0.733&0.755E-04\\
0.30&0.353&0.216&0.245E-01&0.309&-0.739E-01&0.148E-01&0.600&0.105E-03\\
0.20&0.359&0.214&0.182E-01&0.290&-0.639E-01&0.127E-01&0.520&0.149E-03\\
0.10&0.364&0.212&0.154E-01&0.282&-0.591E-01&0.116E-01&0.479&0.143E-03\\
0.01&0.366&0.211&0.145E-01&0.280&-0.577E-01&0.112E-01&0.466&0.133E-03\\
\hline
\multicolumn{9}{c}{FPS}\\
\hline
0.99&0.004&0.332&0.394E-02&0.104&-0.145E-01&0.155E-01&0.180&0.231E-03\\
0.90&0.040&0.320&0.577E-02&0.282&-0.198E-01&0.170E-01&0.226&0.199E-03\\
0.80&0.095&0.302&0.916E-02&0.344&-0.286E-01&0.187E-01&0.299&0.161E-03\\
0.70&0.166&0.278&0.152E-01&0.357&-0.424E-01&0.199E-01&0.406&0.114E-03\\
0.60&0.238&0.254&0.256E-01&0.351&-0.623E-01&0.202E-01&0.557&0.748E-04\\
0.50&0.298&0.234&0.353E-01&0.348&-0.809E-01&0.187E-01&0.682&0.632E-04\\
0.40&0.361&0.213&0.277E-01&0.338&-0.763E-01&0.156E-01&0.618&0.678E-04\\
0.30&0.373&0.209&0.163E-01&0.292&-0.585E-01&0.120E-01&0.482&0.105E-03\\
0.20&0.381&0.206&0.121E-01&0.277&-0.509E-01&0.102E-01&0.418&0.129E-03\\
0.10&0.386&0.205&0.102E-01&0.271&-0.471E-01&0.931E-02&0.385&0.139E-03\\
0.01&0.388&0.204&0.966E-02&0.270&-0.459E-01&0.903E-02&0.374&0.145E-03\\
\hline
\multicolumn{9}{c}{Shen}\\
\hline
0.99&0.002&0.332&0.161&0.051&-0.132&0.701E-01&1.65&0.218E-02\\
0.90&0.027&0.324&0.149&0.157&-0.134&0.645E-01&1.58&0.207E-02\\
0.80&0.059&0.313&0.137&0.220&-0.137&0.580E-01&1.50&0.197E-02\\
0.70&0.096&0.301&0.127&0.266&-0.141&0.510E-01&1.43&0.185E-02\\
0.60&0.138&0.287&0.120&0.301&-0.147&0.436E-01&1.38&0.169E-02\\
0.50&0.185&0.271&0.111&0.328&-0.151&0.365E-01&1.32&0.156E-02\\
0.40&0.234&0.255&0.091&0.345&-0.146&0.308E-01&1.20&0.140E-02\\
0.30&0.279&0.240&0.070&0.348&-0.134&0.262E-01&1.05&0.135E-02\\
0.20&0.303&0.232&0.054&0.338&-0.120&0.230E-01&0.932&0.144E-02\\
0.10&0.305&0.231&0.048&0.325&-0.113&0.215E-01&0.878&0.146E-02\\
0.01&0.306&0.231&0.046&0.322&-0.110&0.210E-01&0.860&0.148E-02\\
\hline
\multicolumn{9}{c}{LS}\\
\hline
0.99&0.003&0.331&0.100E+00&0.054&-0.010&0.570E-01&1.25&0.357E-02\\
0.90&0.031&0.322&0.965E-01&0.166&-0.104&0.532E-01&1.22&0.313E-02\\
0.80&0.067&0.310&0.931E-01&0.229&-0.110&0.485E-01&1.19&0.295E-02\\
0.70&0.108&0.296&0.910E-01&0.273&-0.117&0.434E-01&1.17&0.277E-02\\
0.60&0.154&0.281&0.909E-01&0.307&-0.126&0.377E-01&1.18&0.221E-02\\
0.50&0.203&0.265&0.887E-01&0.331&-0.134&0.320E-01&1.16&0.209E-02\\
0.40&0.255&0.248&0.738E-01&0.345&-0.130&0.270E-01&1.07&0.184E-02\\
0.30&0.299&0.233&0.553E-01&0.343&-0.118&0.229E-01&0.928&0.221E-02\\
0.20&0.316&0.227&0.428E-01&0.327&-0.105&0.201E-01&0.822&0.202E-02\\
0.10&0.317&0.227&0.372E-01&0.314&-0.098&0.187E-01&0.769&0.213E-02\\
0.01&0.318&0.227&0.355E-01&0.311&-0.096&0.182E-01&0.751&0.191E-02\\
\hline
\end{tabular}
\end{minipage}
\end{table*}
\subsubsection{Rotating Magnetized Configurations}
Tables~\ref{real-rot-mix} and \ref{real-mag-mix} 
are devoted to the magnetized sequences with rotation for the four 
kinds of EOSs.
By setting $a=16$ and $\hat{\mu}=0.2$ in Table~\ref{real-rot-mix}, 
$a=16$ and $\hat{\mu}=0.3$ in Table~\ref{real-mag-mix}, 
respectively, we can compute the equilibrium configurations with
comparable toroidal and poloidal fields.
By changing $\hat{\mu}$ in two values, 
we can see clearly the effects of stronger magnetic field.

From the tables, it can be seen that 
there exist maximum ($q_{max}$) and minimum ($q_{min}$) values 
in the axis ratio $q$. This feature does not appear in the sequence
only with the magnetic fields (see Subsection \ref{SMC}), in which
 we can find the physical solutions for any values of $q$
(see Table~\ref{real-non-rot-mix}).

Equilibrium configurations which have larger axis ratio than 
$q_{max}$ have $\hat{\Omega}^2 < 0$ and such configurations are of course 
unphysical. Too strong magnetic field causes 
such a configuration, 
in which ''anti''-centrifugal force is needed. 
On the other hand, configurations with smaller $q$ than $q_{min}$ belong to 
following two types: (1) Due to too strong Lorentz forces, 
the converged solutions have $\hat{\Omega}^2<0$ as explained above. 
(2) the mass of the star sheds from the equatorial surface because the 
centrifugal forces are too strong (MS in the table indicates the
so-called mass-shedding).
Following \cite{tom,yos,yos2}, we call
the sequence ending in the former and the latter type as
magnetic-field-dominated ({\bf MD}) and rotation-dominated ({\bf RD}) sequence, respectively.  
It is noted that larger values of $\hat{\mu}$ make $q_{max}$ smaller 
(compare Table \ref{real-rot-mix} 
with \ref{real-mag-mix}) because $q_{max}$ is determined by the magnetic field 
strength. 

So, all the sequences shown in Table~\ref{real-rot-mix} belong to
the RD sequence regardless of the EOSs, but not for the sequences in 
 Table~\ref{real-mag-mix}. The sequences with SLy or FPS EOSs belong to RD type
and those with Shen or LS EOSs belong to MD type.
As like this, even if we set the same parameter of $a$ and
$\hat{\mu}$, it is found that the type 
of the sequence depends on the EOSs.
 
\begin{figure*}
  \begin{center}
   \vspace*{40pt}
    \begin{tabular}{cc}
      \resizebox{80mm}{!}{\includegraphics{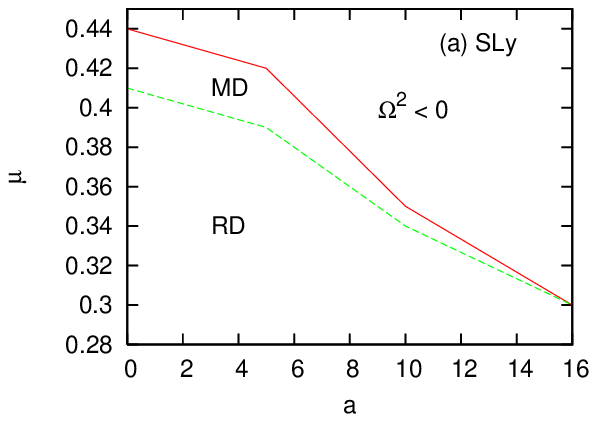}} &
      \resizebox{80mm}{!}{\includegraphics{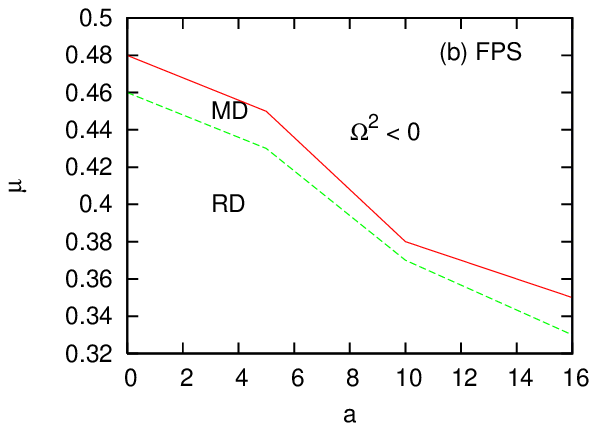}} \\
      \resizebox{80mm}{!}{\includegraphics{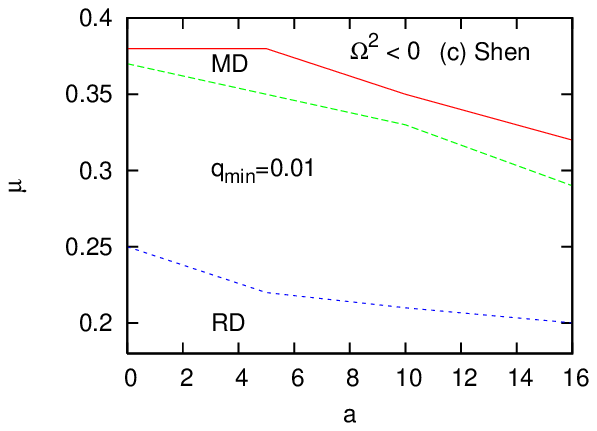}} &
      \resizebox{80mm}{!}{\includegraphics{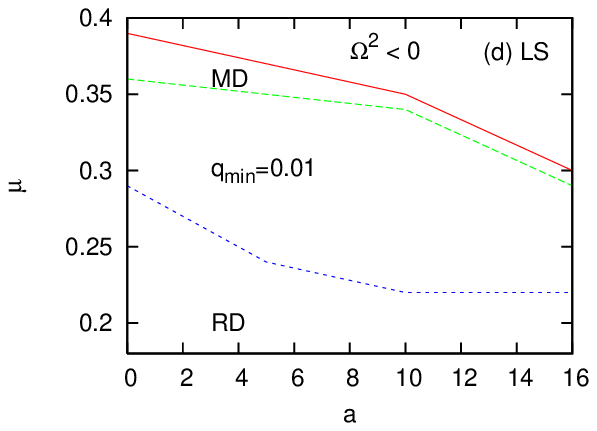}} \\
    \end{tabular}
    \caption{Phase diagrams on $a-\hat{\mu}$ plane for (a)SLy (b) FPS 
   (c) Shen (d) LS. 
    MD and RD 
    mean magnetic-field dominated sequences and rotation-dominated 
    sequences, respectively. $q_{min} = 0.01 $ and $\hat{\Omega}^2<0$ 
    also indicate regions in which nearly toroidal configurations exist and 
    one in which only $\hat{\Omega}^2<0$ solutions exist, respectively.
    \label{Phase}}
  \end{center}
\end{figure*}

In order to see this feature more clearly, we perform a parameter 
search in $a-\hat{\mu}$ parameter space and classify the sequences. 
In Figure~\ref{Phase}, we show a set of phase diagrams 
on $a-\hat{\mu}$ plane with different EOSs. In the figure, the region with 
$\hat{\Omega}^2<0$ shows that we cannot 
obtain any solutions except one with negative angular velocity. 
 ``$q_{min} = 0.01$'' indicates a parameter region in which 
nearly toroidal configurations exist. So we find the equilibrium
sequences with SLy or FPS EOSs are classified into either RD or MD type.
However, the sequences with Shen or 
LS has another type of sequence, in which nearly toroidal 
configurations $q \sim 0 $ exits as in the case of magnetized configurations
without rotation (see Section \ref{SMC}). 
Such a configuration never appears in the model with SLy or FPS 
EOSs.  Looking at the stiffness of EOSs
 in Figure~\ref{adindex} again, it can be seen that the
Shen and LS EOSs are stiffer than SLy and FPS EOSs near the central density adopted here 
($\rho_{max} = 3 \times 10^{14} \rm{g}~\rm{cm}^{-3}$).
Thus the sequence with a nearly toroidal configuration is found to 
appear for the stiffer EOSs. This qualitative feature was 
also noticed in the polytropic studies~\citep{tom,yos,yos2}.

\begin{figure*}
  \begin{center}
  \vspace*{40pt}
    \begin{tabular}{cc}
      \resizebox{80mm}{!}{\includegraphics{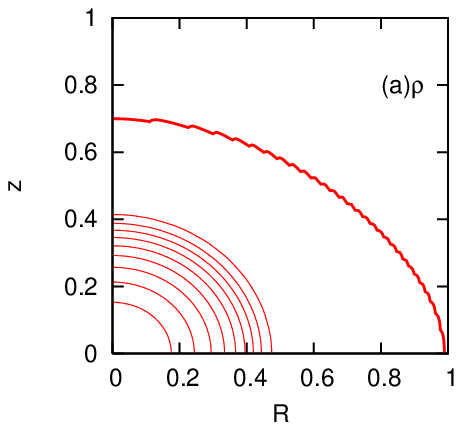}}  &
      \resizebox{80mm}{!}{\includegraphics{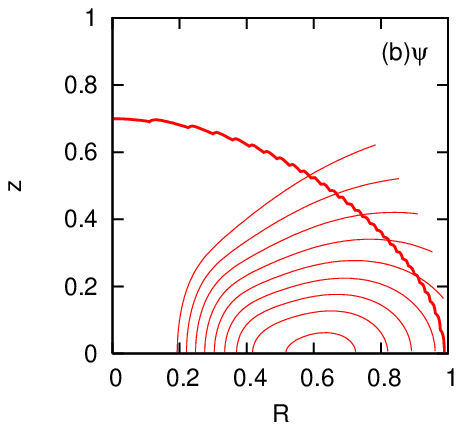}} \\
      \resizebox{80mm}{!}{\includegraphics{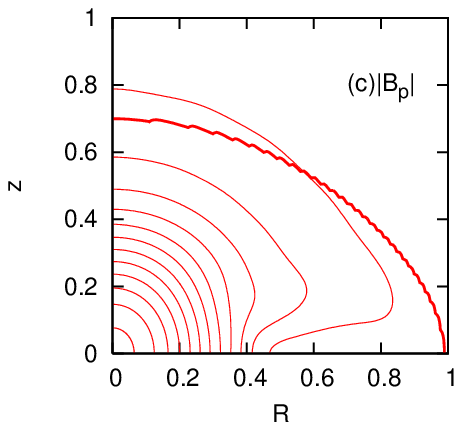}} &
      \resizebox{80mm}{!}{\includegraphics{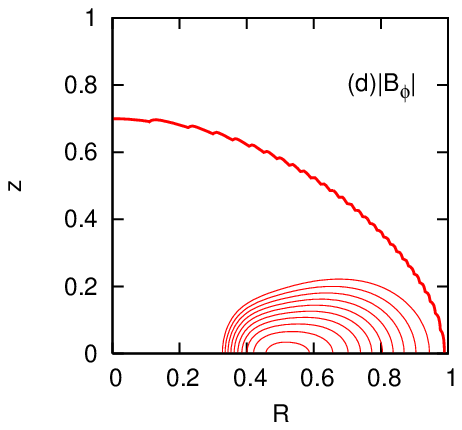}}   \\
    \end{tabular}
    \caption{(a) Density contour, (b) magnetic field lines, 
(c) poloidal component, (d)
toroidal component of magnetic field, in the meridional 
plane for the parameters, $q=0.7$, $a=16$, $\hat{\mu}=0.2$, and SLy EOS. 
The thick, eccentric quarter-ellipse denotes the stellar surface. The 
contours are linearly spaced, i.e., the difference of the physical 
quantities between two adjacent contours in a tenth of the difference 
between the maximum and minimum values.\label{SLyNew}}
  \end{center}
\end{figure*}

\begin{figure*}
  \begin{center}
  \vspace*{40pt}
    \begin{tabular}{cc}
      \resizebox{80mm}{!}{\includegraphics{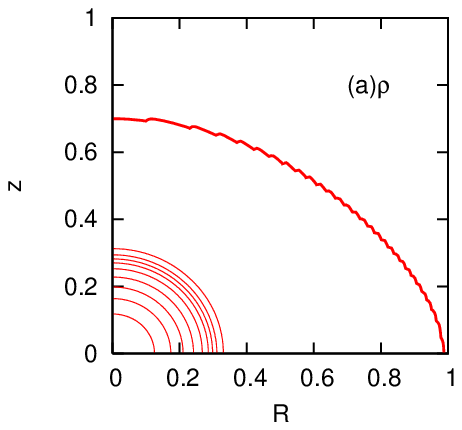}}  &
      \resizebox{80mm}{!}{\includegraphics{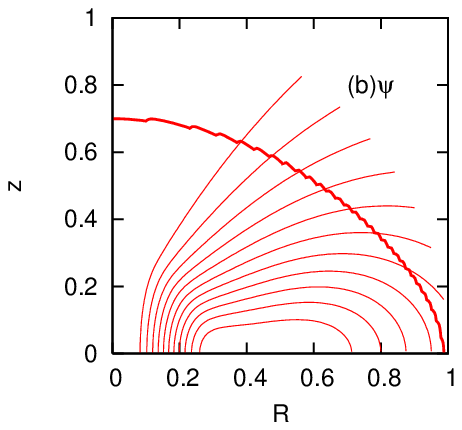}} \\
      \resizebox{80mm}{!}{\includegraphics{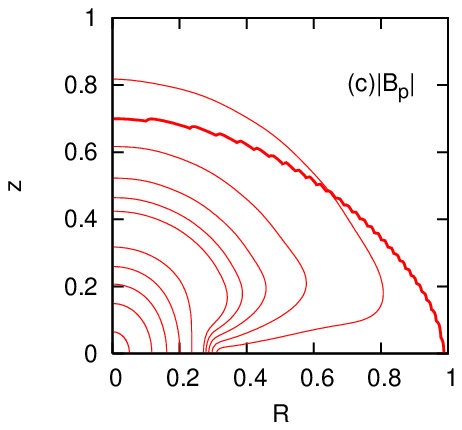}} &
      \resizebox{80mm}{!}{\includegraphics{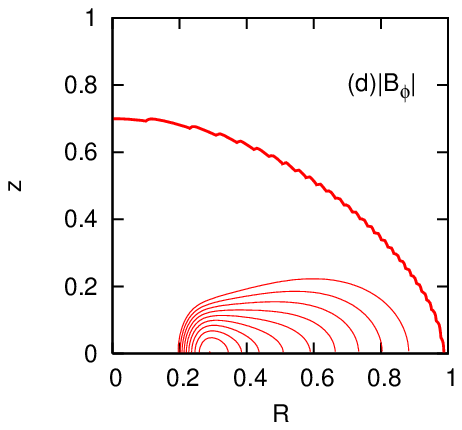}} \\
    \end{tabular}
    \caption{Same as Figure~\ref{SLyNew} but for FPS EOS.\label{FPSNew}}
  \end{center}
\end{figure*}

\begin{figure*}
  \begin{center}
  \vspace*{40pt}
    \begin{tabular}{cc}
      \resizebox{80mm}{!}{\includegraphics{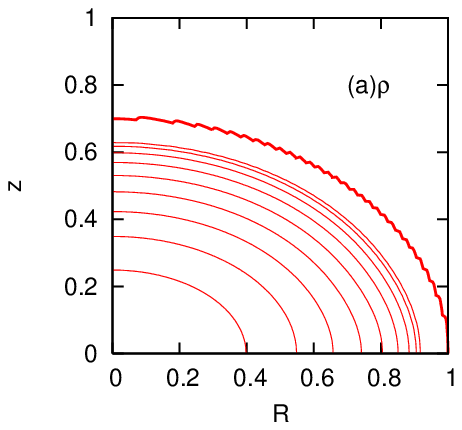}} &
      \resizebox{80mm}{!}{\includegraphics{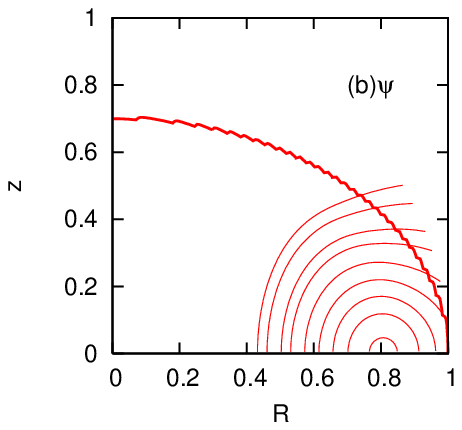}} \\
      \resizebox{80mm}{!}{\includegraphics{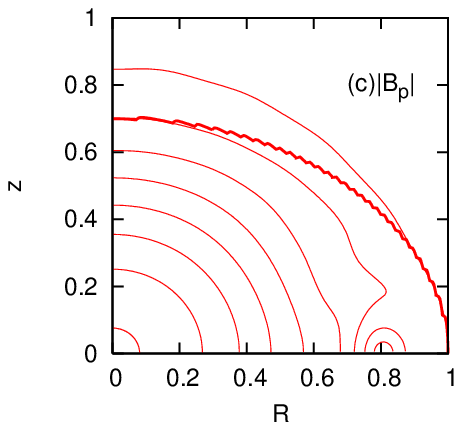}} &
      \resizebox{80mm}{!}{\includegraphics{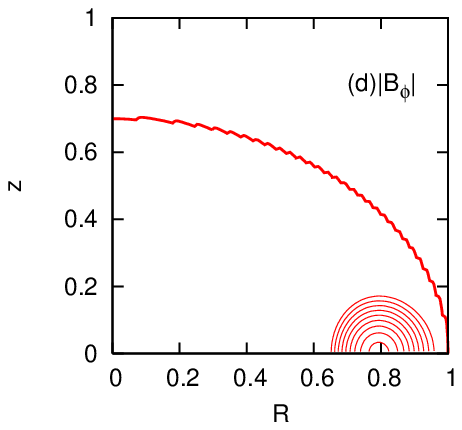}}   \\
    \end{tabular}
    \caption{Same as Figure~\ref{SLyNew} but for Shen EOS.\label{ShenNew}}
  \end{center}
\end{figure*}

\begin{figure*}
  \begin{center}
  \vspace*{40pt}
    \begin{tabular}{cc}
      \resizebox{80mm}{!}{\includegraphics{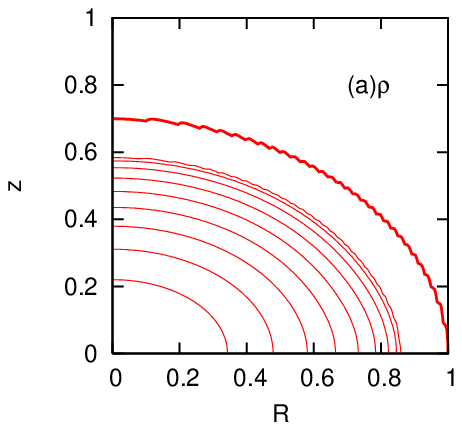}}  &
      \resizebox{80mm}{!}{\includegraphics{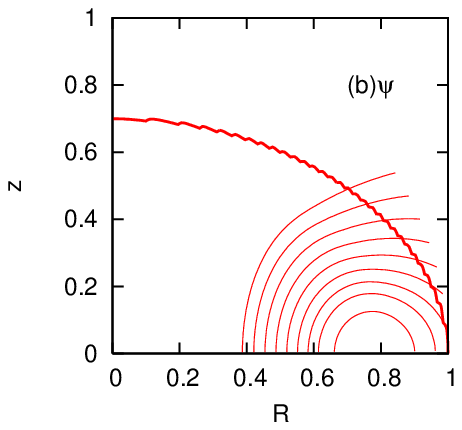}} \\
      \resizebox{80mm}{!}{\includegraphics{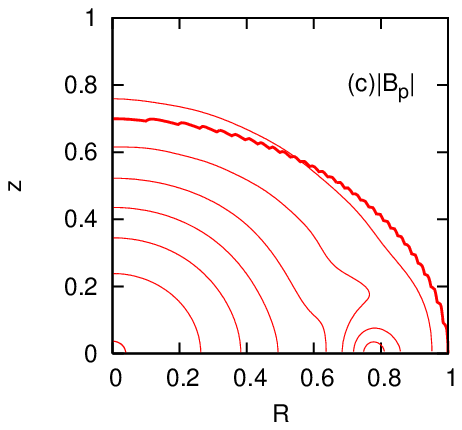}} &
      \resizebox{80mm}{!}{\includegraphics{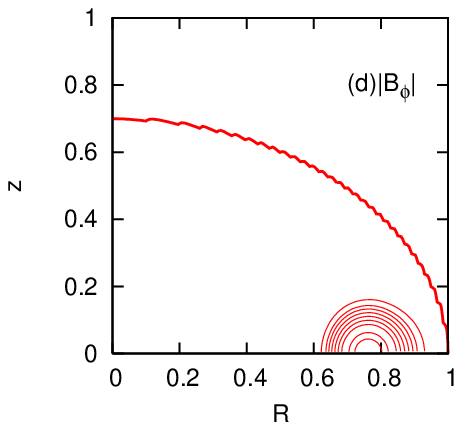}}   \\
    \end{tabular}
    \caption{Same as Figure~\ref{SLyNew} but for LS EOS.
\label{LSNew}}
  \end{center}
\end{figure*}

Then we move on to investigate the structures of the 
equilibrium configurations.
Figures~\ref{SLyNew}-\ref{LSNew} show the distributions of 
density, and toroidal/poloidal magnetic fields of models 
characterized by $q=0.7$, $a=16$, $\hat{\mu}=0.2$ for SLy, FPS, Shen, and 
LS EOSs, respectively. The reason why these values of the parameters are chosen is 
that we want to fix $q$ for the different EOS models. For example, 
there does not exist the common value of $q$ if we set the value of
$\hat{\mu}$ as $0.3$ (see Table~\ref{real-mag-mix}). 
$q=0.7$ taken here is the smallest common value in the sequences with $\hat{\mu}=0.2$ 
(see Table~\ref{real-rot-mix}).

It is seen that the density distributions of stars with SLy or FPS EOSs are
more concentrated at the center than those of the stars with Shen or 
LS EOSs. These feature are due to the stiffness 
of EOSs as mentioned above. We also find the toroidal fields
for SLy or FPS EOSs distribute in relatively wider regions in the
vicinity of the equatorial plane than for
 Shen or LS EOSs, in which the toroidal fields concentrate near the stellar surfaces. 
It is noted that the toroidal fields only exist in the
interior of the stars, which is due to the choice of the functional form of $\kappa$ 
(see equation (\ref{eq27})). 
Near the rotational axis, the poloidal magnetic field behaves like a
dipole field, where the 
toroidal fields are weak. However in the vicinity of the
equatorial plane where the toroidal fields become strong, the poloidal
fields become distorted. 
The region where the magnetic field is mixed depends 
on the EOSs. The stars with the softer EOS (SLy or FPS) tend to have wider 
mixed region than those with the stiffer EOS (Shen or LS). 
The magnetic field lines $\Psi$ may be good a tool to
see the structures, showing the tori of twisted field lines 
around the symmetry axis inside the star and the untwisted poloidal 
field lines, which penetrate the stellar surface to continue to the
vacuum. It is noted that this universality of the twisted-torus
structures of the magnetic fields was seen also in the polytropic stars
\citep{yos2}. Therefore our results for the realistic EOSs 
may be regarded as a further generalization of their results.
\begin{table*}
\centering
 \begin{minipage}{140mm}
\caption{Physical quantities for rotating magnetized stars with $a=16$
 and $\hat{\mu}= 0.2$.\label{real-rot-mix}}
\begin{tabular}{ccccccccccc}
\hline\hline
q & $H/|W|$ & $U/|W|$ & $T/|W|$ & $|\hat{W}|$ &$\hat{\Omega}^2$ & $\hat{C}$ & 
$\hat{\beta}$ & $\hat{M}$ & h & VC \\
\hline
\multicolumn{11}{c}{SLy}\\
\hline
$\hat{\Omega}^2<0$&-&-&-&-&-&-&-&-&-\\
0.90&0.350E-1&0.321&0.855E-3&0.147E-1&0.536E-3&-0.347E-1&0.249E-1&0.399&0.320&0.128E-3\\
0.80&0.283E-1&0.316&0.117E-1&0.898E-2&0.720E-2&-0.290E-1&0.202E-1&0.297&0.270&0.167E-3\\
0.70&0.198E-1&0.315&0.176E-1&0.452E-2&0.108E-1&-0.220E-1&0.154E-1&0.197&0.210&0.216E-3\\
0.63&0.146E-1&0.316&0.184E-1&0.251E-2&0.114E-1&-0.172E-1&0.122E-1&0.138&0.199&0.261E-3\\
MS&-&-&-&-&-&-&-&-&-\\
\hline
\multicolumn{11}{c}{FPS}\\
\hline
$\hat{\Omega}^2<0$&-&-&-&-&-&-&-&-&-\\
0.94&0.168E-1&0.328&0.374E-3&0.417E-2&0.247E-3&-0.157E-1&0.156E-1&0.187&0.189&0.225E-3\\
0.90&0.152E-1&0.327&0.234E-2&0.341E-2&0.154E-2&-0.146E-1&0.143E-1&0.166&0.180&0.242E-3\\
0.80&0.113E-1&0.326&0.583E-2&0.191E-2&0.383E-2&-0.116E-1&0.114E-1&0.117&0.166&0.306E-3\\
0.70&0.783E-2&0.326&0.729E-2&0.941E-3&0.480E-2&-0.866E-2&0.858E-2&0.765E-1&0.160&0.402E-3\\
0.65&0.644E-2&0.326&0.742E-2&0.632E-3&0.489E-2&-0.735E-2&0.733E-2&0.603E-1&0.168&0.477E-3\\
MS&-&-&-&-&-&-&-&-&-\\
\hline
\multicolumn{11}{c}{Shen}\\
\hline
$\hat{\Omega}^2<0$&-&-&-&-&-&-&-&-&-\\
0.80&0.605E-1&0.311&0.216E-2&0.132&0.138E-2&-0.134&0.573E-1&0.146E+1&0.424&0.197E-2\\
0.70&0.672E-1&0.289&0.315E-1&0.106&0.183E-1&-0.129&0.485E-1&0.129E+1&0.458&0.182E-2\\
0.60&0.756E-1&0.265&0.644E-1&0.814E-1&0.335E-1&-0.122&0.397E-1&0.111E+1&0.497&0.166E-2\\
0.50&0.860E-1&0.237&0.101&0.596E-1&0.461E-1&-0.113&0.309E-1&0.929&0.537&0.148E-2\\
0.40&0.991E-1&0.204&0.144&0.406E-1&0.550E-1&-0.101&0.224E-1&0.750&0.582&0.124E-2\\
MS&-&-&-&-&-&-&-&-&-\\
\hline
\multicolumn{11}{c}{LS}\\
\hline
$\hat{\Omega}^2<0$&-&-&-&-&-&-&-&-&-\\
0.81&0.669E-1&0.309&0.143E-2&0.881E-1&0.883E-3&-0.106&0.482E-1&0.115E+1&0.502&0.289E-2\\
0.80&0.675E-1&0.307&0.400E-2&0.862E-1&0.246E-2&-0.106&0.475E-1&0.114E+1&0.505&0.291E-2\\
0.70&0.749E-1&0.287&0.313E-1&0.675E-1&0.176E-1&-0.101&0.401E-1&0.988&0.535&0.275E-2\\
0.60&0.834E-1&0.264&0.615E-1&0.495E-1&0.312E-1&-0.934E-1&0.326E-1&0.826&0.565&0.242E-2\\
0.50&0.905E-1&0.240&0.929E-1&0.304E-1&0.422E-1&-0.799E-1&0.246E-1&0.622&0.584&0.238E-2\\
0.48&0.865E-1&0.239&0.972E-1&0.245E-1&0.439E-1&-0.732E-1&0.226E-1&0.547&0.567&0.236E-2\\
MS&-&-&-&-&-&-&-&-&-\\
\hline
\end{tabular}
\end{minipage}
\end{table*}
\begin{table*}
\centering
 \begin{minipage}{140mm}
\caption{Same as Table~\ref{real-rot-mix}, but with $a=16$
 and $\hat{\mu}= 0.3$.\label{real-mag-mix}}
\begin{tabular}{ccccccccccc}
\hline\hline
q & $H/|W|$ & $U/|W|$ & $T/|W|$ & $|\hat{W}|$ &$\hat{\Omega}^2$ & $\hat{C}$ & 
$\hat{\beta}$ & $\hat{M}$ & h & VC \\
\hline
\multicolumn{11}{c}{SLy}\\
\hline
$\hat{\Omega}^2<0$&-&-&-&-&-&-&-&-&-&-\\
0.70&0.147&0.282&0.283E-2&0.195E-1&0.144E-2&-0.494E-1&0.232E-1&0.474&0.471&0.820E-4\\
0.60&0.425E-1&0.305&0.214E-1&0.323E-2&0.125E-1&-0.206E-1&0.128E-1&0.161&0.203&0.241E-3\\
MS&-&-&-&-&-&-&-&-&-&-\\
\hline
\multicolumn{11}{c}{FPS}\\
\hline
$\hat{\Omega}^2<0$&-&-&-&-&-&-&-&-&-&-\\
0.86&0.492E-1&0.317&0.301E-3&0.525E-2&0.187E-3&-0.195E-1&0.162E-1&0.214&0.217&0.207E-3\\
0.80&0.372E-1&0.318&0.483E-2&0.326E-2&0.304E-2&-0.158E-1&0.135E-1&0.161&0.179&0.255E-3\\
0.70&0.226E-1&0.321&0.793E-2&0.135E-2&0.507E-2&-0.107E-1&0.967E-2&0.950E-1&0.159&0.361E-3\\
0.63&0.161E-1&0.323&0.819E-2&0.700E-3&0.529E-2&-0.802E-2&0.750E-2&0.640E-1&0.167&0.459E-3\\
MS&-&-&-&-&-&-&-&-&-&-\\
\hline
\multicolumn{11}{c}{Shen}\\
\hline
$\hat{\Omega}^2<0$&-&-&-&-&-&-&-&-&-&-\\
0.52&0.190&0.269&0.518E-3&0.106&0.254E-3&-0.146&0.367E-1&0.129E+01&0.491&0.153E-2\\
0.50&0.195&0.264&0.635E-2&0.103&0.304E-2&-0.146&0.350E-1&0.127E+01&0.496&0.154E-2\\
0.40&0.222&0.237&0.332E-1&0.850E-1&0.136E-1&-0.142&0.275E-1&0.115E+01&0.525&0.135E-2\\
0.30&0.260&0.216&0.459E-1&0.654E-1&0.151E-1&-0.131&0.222E-1&0.101E+01&0.565&0.124E-2\\
0.20&0.305&0.213&0.273E-1&0.505E-1&0.713E-2&-0.117&0.201E-1&0.897&0.610&0.139E-2\\
0.11&0.331&0.222&0.110E-2&0.457E-1&0.259E-3&-0.110&0.200E-1&0.858&0.615&0.148E-2\\
$\hat{\Omega}^2<0$&-&-&-&-&-&-&-&-&-&-\\
\hline
\multicolumn{11}{c}{LS}\\
\hline
$\hat{\Omega}^2<0$&-&-&-&-&-&-&-&-&-&-\\
0.51&0.217&0.259&0.243E-2&0.813E-1&0.112E-2&-0.128&0.311E-1&0.111E+01&0.569&0.205E-2\\
0.50&0.220&0.256&0.512E-2&0.805E-1&0.233E-2&-0.128&0.304E-1&0.110E+01&0.571&0.194E-2\\
0.40&0.251&0.230&0.289E-1&0.695E-1&0.113E-1&-0.127&0.241E-1&0.102E+01&0.600&0.171E-2\\
0.30&0.298&0.212&0.322E-1&0.533E-1&0.981E-2&-0.117&0.199E-1&0.905&0.648&0.191E-2\\
0.21&0.350&0.216&0.842E-3&0.418E-1&0.204E-3&-0.104&0.186E-1&0.811&0.690&0.105E-2\\
$\hat{\Omega}^2<0$&-&-&-&-&-&-&-&-&-&-\\
\hline
\end{tabular}
\end{minipage}
\end{table*}
\subsection{General Relativistic Case}\label{grresult}
As mentioned in Section \ref{intro}, when one tries to perform
a qualitative investigation on the equilibrium configurations of 
magnetized rotating stars, the general 
relativistic effect must be taken into account.
Since a full treatment is 
beyond scope of our paper,
 we here try to include the effect by adding a spherically
 general relativistic correction to the Newtonian gravitational
 potential. Although this method is similar to the approach reported
 in \cite{rampp,mar}, the definition of general relativistic potential and 
its boundary condition are modified to be appropriate for this study. 
 Details of our approach and its verification are shown in appendix 
\ref{GR correct}. In this subsection, we report the models with the correction.
\subsubsection{Basic property}
In Table~\ref{gr1}, physical quantities of magnetized 
stars, such as magnetic field at pole $B_p$, one at center $B_c$, 
baryon mass $M$, rotation period $P$, and radius $R$, are shown 
with $a=20$ and $\hat{\mu}=0.1$. Maximum density is taken to be 
$10^{15}\rm{g}~\rm{cm}^{-3}$, typically 
where the general relativistic correction cannot be negligible. 
Since our treatment for the correction may break 
down if the equilibrium star is highly deformed, we only focus on the 
equilibrium sequences with mildly strong magnetic field with the
comparable strength of poloidal and toroidal fields (see $h$ in
Table~\ref{gr1}). 

From the table, it is found that there exist 
$q_{max}$ and $q_{min}$ as same in the Newtonian sequences. 
We find that all the sequences belong to the
rotation-dominated type, irrespective of EOSs,
with the maximum and minimum value of axis ratio 
being almost the same, typically $q_{max} \sim 0.9 $ and $q_{min} \sim 0.6 $. 
More interestingly, we find the GR effect does not change the basic 
property of the magnetized stars drastically. 

The values of the magnetic field at pole and center are 
also comparable, typically $B_p \sim 10^{16} [\rm{G}] $ and $B_c \sim 10^{17} 
[\rm{G}]$.
The baryon masses of these stars are in the range between 
$1.7 M_\odot $ - $3.6 M_\odot$ and the stars with FPS EOS tend to have 
lighter masses and those with Shen EOS.
Masses of the stars with SLy and LS EOSs are shown to be comparable.
Comparing with the observations of magnetars \citep{kapsi}, the obtained magnetic
fields are a bit larger and the rotation periods are an order of milliseconds, which
 are too rapid in comparison with the observed ones of seconds. 
We discuss this point in the Subsection \ref{sec:proto-mag}.

\begin{figure*}
  \begin{center}
  \vspace*{174pt}
    \begin{tabular}{cc}
      \resizebox{80mm}{!}{\includegraphics{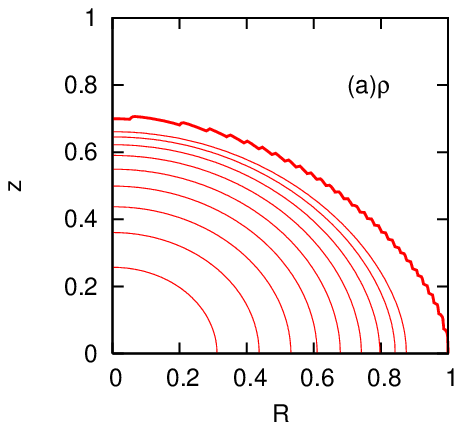}} &
      \resizebox{80mm}{!}{\includegraphics{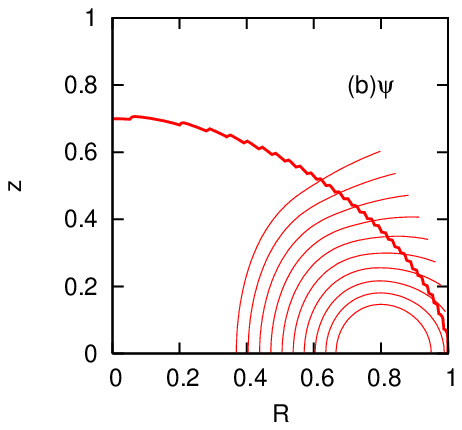}} \\
      \resizebox{80mm}{!}{\includegraphics{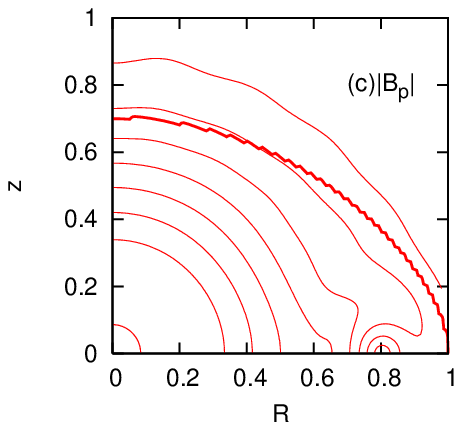}} &  
      \resizebox{80mm}{!}{\includegraphics{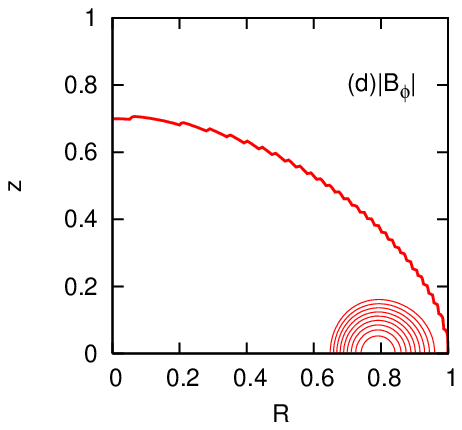}} \\
    \end{tabular}
    \caption{(a) Density contour, (b) magnetic field lines, (c) poloidal 
component of magnetic field, (d) 
toroidal component of magnetic field, in the meridional 
plane for the parameters, $q=0.7$, $a=20$, $\hat{\mu}=0.1$, and SLy EOS. 
General relativistic correction is included. 
The thick, eccentric quarter-ellipse denotes the stellar surface. The 
contours are linearly spaced, i.e., the difference of the physical 
quantities between two adjacent contours in a tenth of the difference 
between the maximum and minimum values. To make a comparison with the 
Newtonian result easy, these quantities are shown in the non-dimensional unit.
\label{GRSLy}}
  \end{center}
\end{figure*}

\begin{figure*}
  \begin{center}
  \vspace*{174pt}
    \begin{tabular}{cc}
      \resizebox{80mm}{!}{\includegraphics{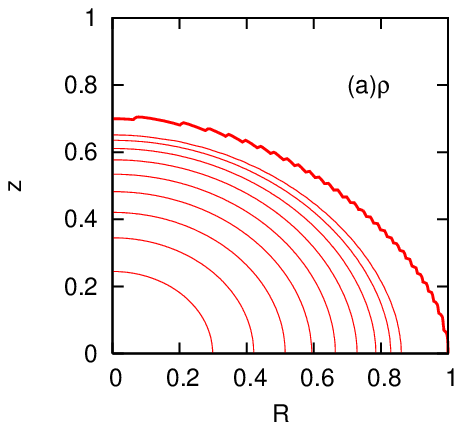}} &
      \resizebox{80mm}{!}{\includegraphics{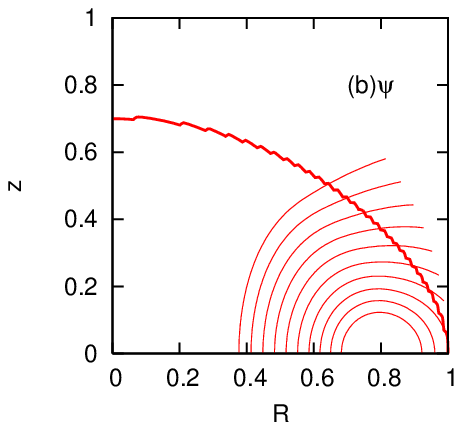}} \\
      \resizebox{80mm}{!}{\includegraphics{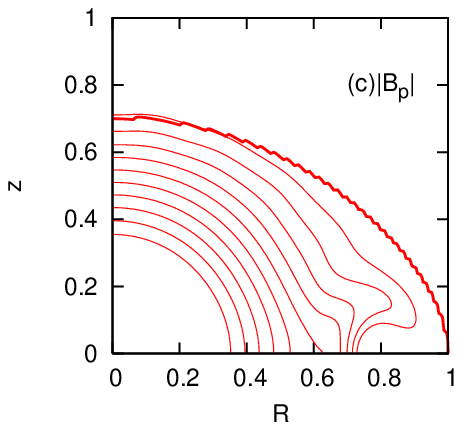}} &  
      \resizebox{80mm}{!}{\includegraphics{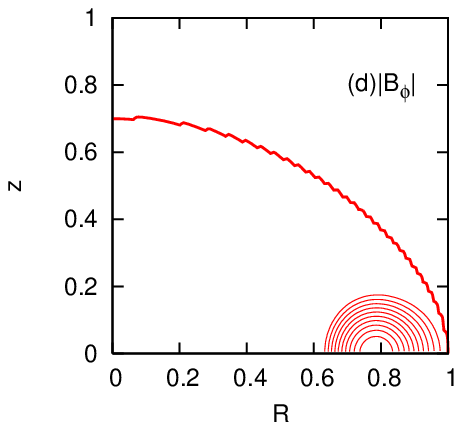}} \\
    \end{tabular}
    \caption{Same as Fig.~\ref{GRSLy} but for FPS EOS.\label{GRFPS}}
  \end{center}
\end{figure*}

\begin{figure*}
  \begin{center}
  \vspace*{174pt}
    \begin{tabular}{cc}
      \resizebox{80mm}{!}{\includegraphics{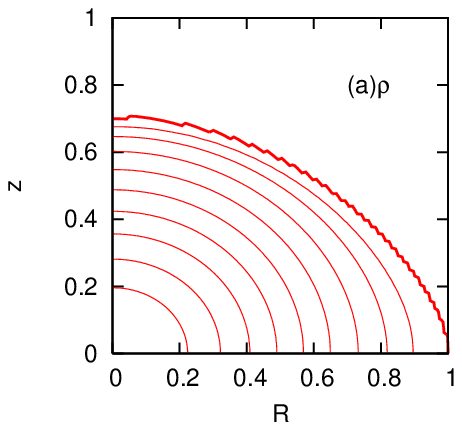}} &
      \resizebox{80mm}{!}{\includegraphics{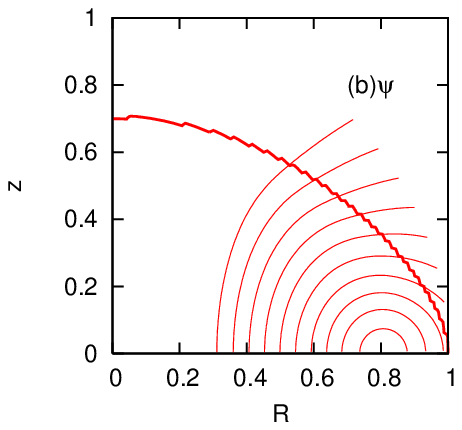}} \\
      \resizebox{80mm}{!}{\includegraphics{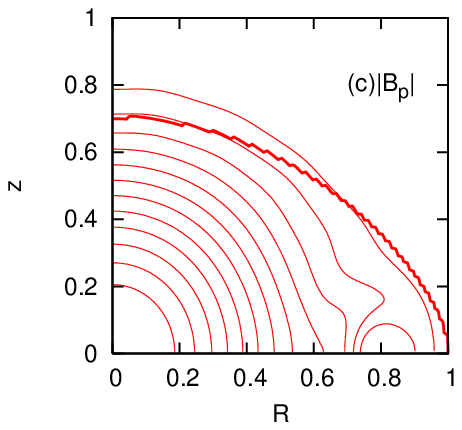}} &  
      \resizebox{80mm}{!}{\includegraphics{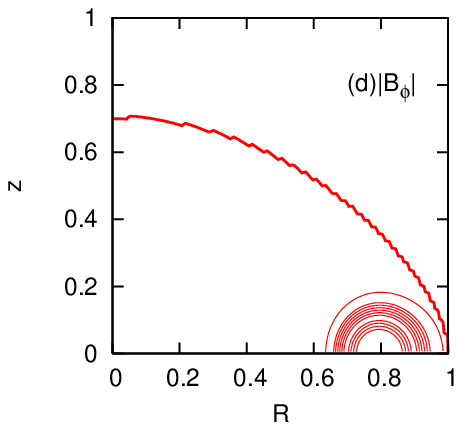}} \\
    \end{tabular}
    \caption{Same as Fig.~\ref{GRSLy} but for Shen EOS.\label{GRShen}}
  \end{center}
\end{figure*}

\begin{figure*}
  \begin{center}
  \vspace*{174pt}
    \begin{tabular}{cc}
      \resizebox{80mm}{!}{\includegraphics{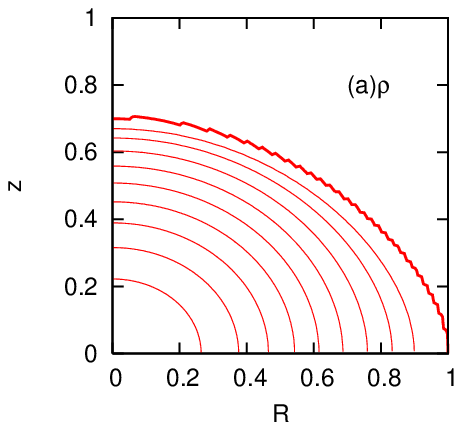}} &
      \resizebox{80mm}{!}{\includegraphics{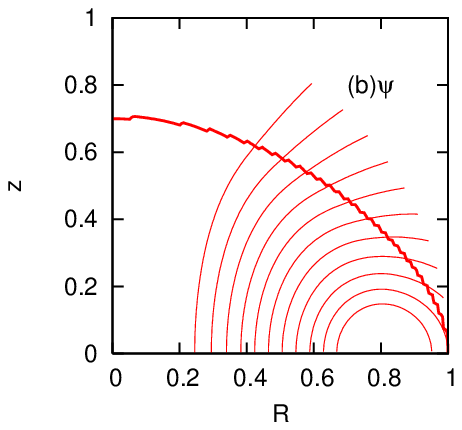}} \\
      \resizebox{80mm}{!}{\includegraphics{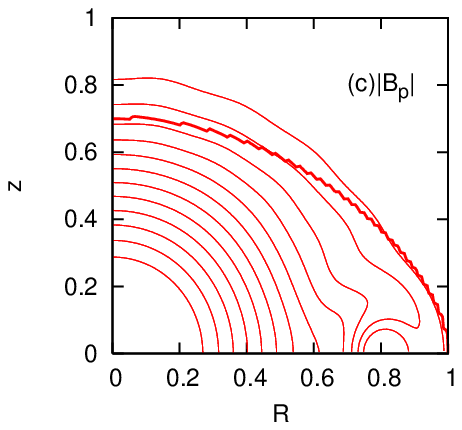}} &  
      \resizebox{80mm}{!}{\includegraphics{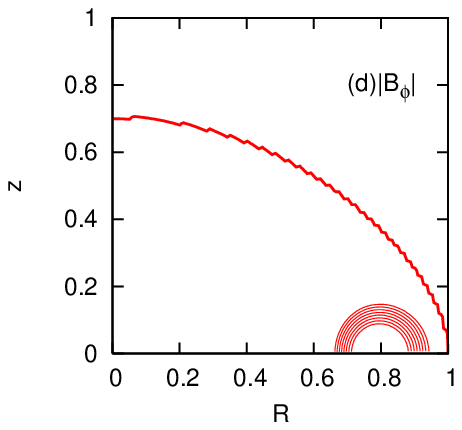}} \\
    \end{tabular}
    \caption{Same as Fig.~\ref{GRSLy} but for LS EOS.\label{GRLS}}
  \end{center}
\end{figure*}

Figures~\ref{GRSLy}, \ref{GRFPS}, \ref{GRShen}, and 
\ref{GRLS} are the density and magnetic field distributions of the equilibrium 
configurations with $q=0.7$ in Table~\ref{gr1}. Note again that this value is 
the smallest common value in the configurations with the different EOSs. 
It can be seen that the configurations of density and the magnetic
fields are rather similar regardless of the EOSs. The toroidal components of 
the magnetic fields concentrate near the stellar surface. The poloidal 
fields are distorted in the region, in which the toroidal fields are strong, and 
its shapes are dipole like near the rotation axis. 
Despite of the incursion of the general relativistic correction, 
the structures of the magnetic field lines are 
twisted tori as same in the Newtonian case.
The reason of these similarity is that the employed EOSs become sufficiently stiff
(Figure~\ref{adindex}) below the maximum density of
$10^{15}\rm{g}~\rm{cm}^{-3}$ 
and that the effects on the configurations become smaller.
 It is seen that Shen and LS EOSs become softer at the density regime due
to the increasing proton fraction. However it is noted that the equilibrium
configurations are affected whether or not 
the EOSs become already stiffened at the smaller density regimes than
 the maximum density. In the two EOSs, 
this indeed occurs at  $\sim 10^{14}\rm{g}~\rm{cm}^{-3}$. After the stiffening, the
effects of the difference of the EOSs on the configurations become
small with increasing the maximum density. 

\begin{table*}
\centering
 \begin{minipage}{140mm}
\caption{Physical quantities for the rotating magnetized stars 
with $a=20$, $\hat{\mu}=0.1$ and general relativistic correction.
Note that $R$ is an equatorial radius.
\label{gr1}} 
\begin{tabular}{ccccccc}
\hline\hline
$q$ & $Bp[\rm{G}]$ & $Bc[\rm{G}]$ & $M[M_\odot]$ & $P[\rm{ms}]$ & $R[\rm{km}]$ & h \\
\hline
\multicolumn{7}{c}{SLy}\\
\hline
$\hat{\Omega}^2<0$&-&-&-&-&-&-\\
0.97&0.663E+17&0.273E+18&0.174E+01&0.336E+01&0.117E+02&0.393E+00\\
0.90&0.723E+17&0.277E+18&0.185E+01&0.120E+01&0.124E+02&0.429E+00\\
0.80&0.826E+17&0.282E+18&0.203E+01&0.831E+00&0.135E+02&0.491E+00\\
0.70&0.913E+17&0.282E+18&0.222E+01&0.711E+00&0.152E+02&0.560E+00\\
0.63&0.908E+17&0.267E+18&0.228E+01&0.684E+00&0.167E+02&0.563E+00\\
MS&-&-&-&-&-&-\\
\hline
\multicolumn{7}{c}{FPS}\\
\hline
$\hat{\Omega}^2<0$&-&-&-&-&-&-\\
0.97&0.581E+17&0.251E+18&0.135E+01&0.464E+01&0.111E+02&0.425E+00\\
0.90&0.635E+17&0.255E+18&0.143E+01&0.136E+01&0.117E+02&0.459E+00\\
0.80&0.732E+17&0.259E+18&0.156E+01&0.924E+00&0.128E+02&0.516E+00\\
0.70&0.813E+17&0.258E+18&0.170E+01&0.783E+00&0.143E+02&0.572E+00\\
0.63&0.818E+17&0.245E+18&0.176E+01&0.748E+00&0.158E+02&0.574E+00\\
MS&-&-&-&-&-&-\\
\hline
\multicolumn{7}{c}{Shen}\\
\hline
$\hat{\Omega}^2<0$&-&-&-&-&-&-\\
0.98&0.670E+17&0.277E+18&0.282E+01&0.372E+01&0.140E+02&0.293E+00\\
0.90&0.732E+17&0.280E+18&0.301E+01&0.106E+01&0.149E+02&0.326E+00\\
0.80&0.816E+17&0.282E+18&0.328E+01&0.761E+00&0.163E+02&0.380E+00\\
0.70&0.876E+17&0.276E+18&0.353E+01&0.668E+00&0.183E+02&0.459E+00\\
0.65&0.855E+17&0.263E+18&0.358E+01&0.656E+00&0.197E+02&0.463E+00\\
MS&-&-&-&-&-&-\\
\hline
\multicolumn{7}{c}{LS}\\
\hline
$\hat{\Omega}^2<0$&-&-&-&-&-&-\\
0.98&0.613E+17&0.259E+18&0.186E+01&0.998E+01&0.123E+02&0.333E+00\\
0.90&0.673E+17&0.263E+18&0.199E+01&0.125E+01&0.131E+02&0.367E+00\\
0.80&0.762E+17&0.266E+18&0.217E+01&0.870E+00&0.143E+02&0.419E+00\\
0.70&0.834E+17&0.263E+18&0.237E+01&0.747E+00&0.160E+02&0.483E+00\\
0.62&0.816E+17&0.245E+18&0.244E+01&0.719E+00&0.180E+02&0.491E+00\\
MS&-&-&-&-&-&-\\
\hline
\end{tabular}
\end{minipage}
\end{table*}
\subsubsection{Relations between physical quantities and maximum density}
\begin{figure*}
  \begin{center}
  \vspace*{174pt}
    \begin{tabular}{cc}
      \resizebox{80mm}{!}{\includegraphics{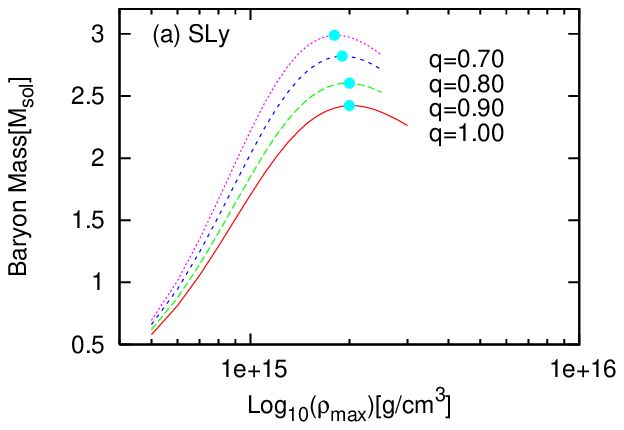}} & 
      \resizebox{80mm}{!}{\includegraphics{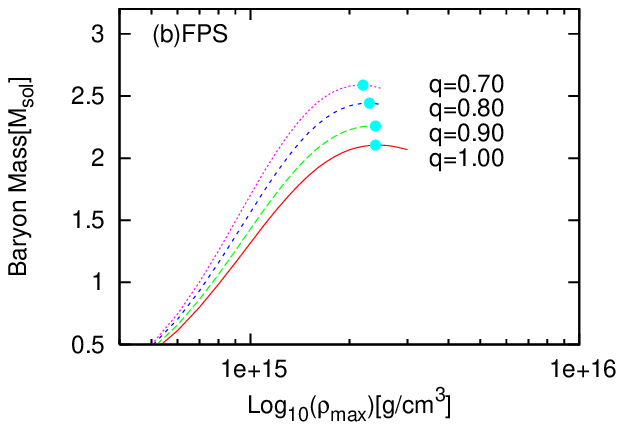}} \\
      \resizebox{80mm}{!}{\includegraphics{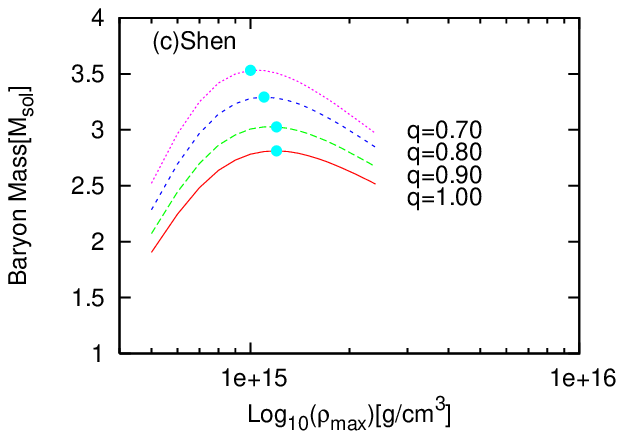}} &
      \resizebox{80mm}{!}{\includegraphics{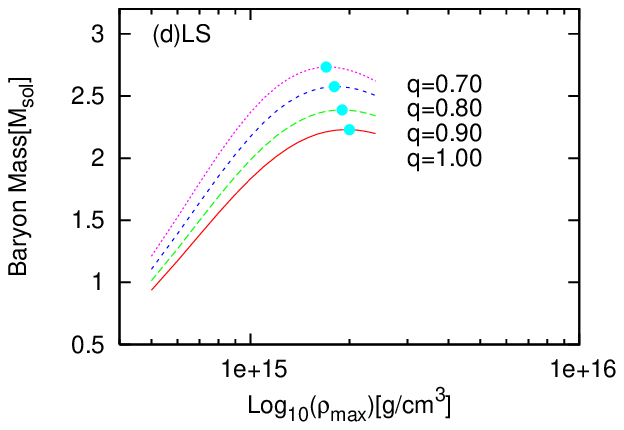}} \\
    \end{tabular}
    \caption{Maximum density-baryon mass relation in the equilibrium 
configurations with (a) SLy, (b) FPS, (c) Shen, and (d) LS 
for the parameter, $a=20$ and $\hat{\mu}=0.1$. As a reference, 
the relation in the spherical configuration are shown as $q=1.0$.
 Solid, long-dashed, 
short-dashed, and dotted lines correspond to axis ratio $q=1.0$, $q=0.9$, 
$q=0.8$, and $q=0.7$, respectively. Closed circles shown in the each figures 
are maximum mass points.\label{rho-mass}}
  \end{center}
\end{figure*}

\begin{figure*}
  \begin{center}
  \vspace*{174pt}
    \begin{tabular}{cc}
      \resizebox{80mm}{!}{\includegraphics{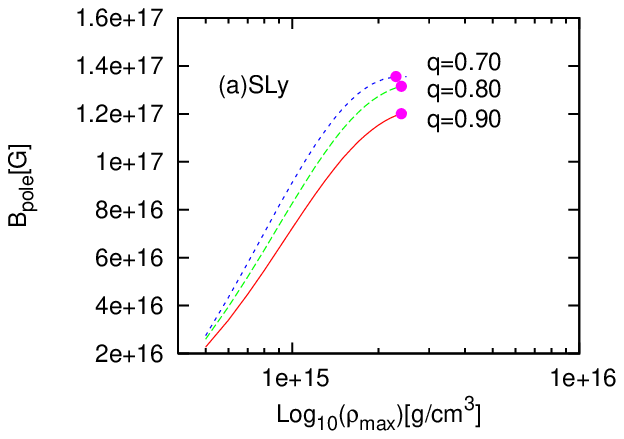}} &
      \resizebox{80mm}{!}{\includegraphics{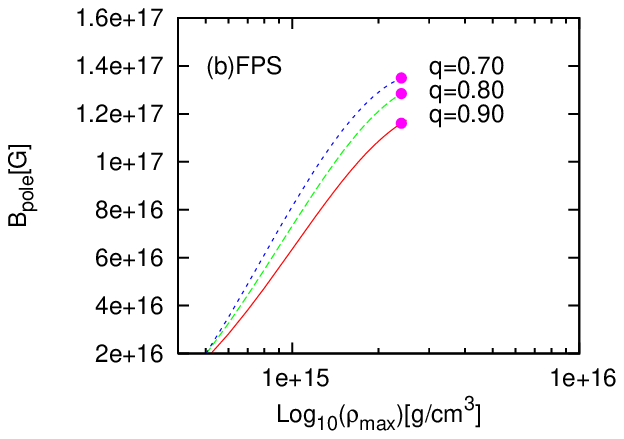}} \\
      \resizebox{80mm}{!}{\includegraphics{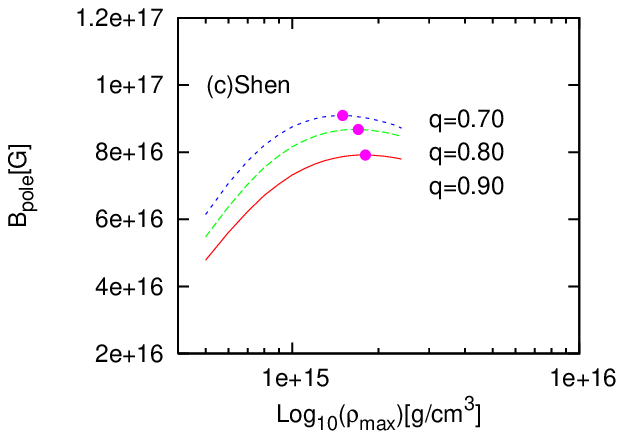}} &
      \resizebox{80mm}{!}{\includegraphics{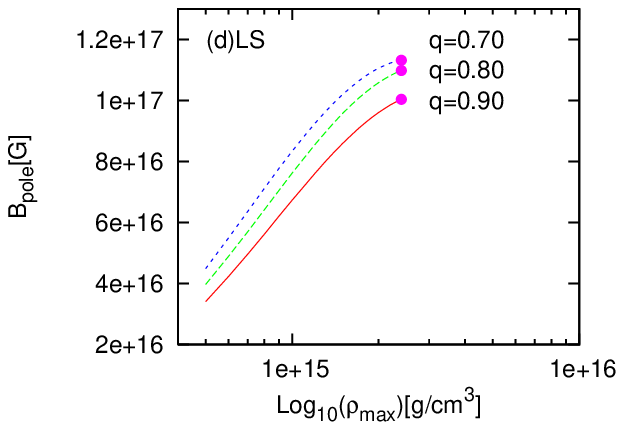}} \\
    \end{tabular}
    \caption{ Magnetic fields strength at pole as functions of the 
maximum densities 
in the same sequences shown in Figure~\ref{rho-mass}.\label{rho-Bp}}
  \end{center}
\end{figure*}

The relation between the physical quantities and the maximum density is an important 
information and it sometimes helps us to understand the stability of 
equilibrium configuration. 
Figure~\ref{rho-mass} shows maximum density and baryon mass relations in 
the magnetized rotating stars for the parameters, $a=12$ and 
$\hat{\mu}=0.1$ for the different EOSs. Long-dashed, 
short-dashed, and dotted lines represent the relations for 
$q=0.90$, $q=0.80$, and $q=0.70$, respectively. As a reference relation, 
we also show one in the spherical configuration as $q=1.0$. 
All sequences 
reach the mass-shedding limit at $q \sim 0.60$. Closed circles correspond to 
the maximum mass points. As the axis ratios become small, the maximum mass points 
shift to left, i.e. lower density region. We find that the increase of the 
maximum mass is up to about twenty percents for the every EOS employed here. 
In Figure \ref{rho-Bp}, the magnetic fields at the pole are shown as a function of
the maximum densities in the same sequences. For the sequences 
with SLy, FPS, and LS EOSs, the magnetic field at pole 
is an increasing function of the maximum density for any axis ratio. 
On the other hand, it is found that 
only for the sequences with Shen EOS, these exist maximum of the polar 
magnetic fields around $\rho 
\sim 2 \times 10^{15} \rm{g}~\rm{cm}^{-3}$.

\begin{figure*}
  \begin{center}
  \vspace*{174pt}
    \begin{tabular}{cc}
      \resizebox{80mm}{!}{\includegraphics{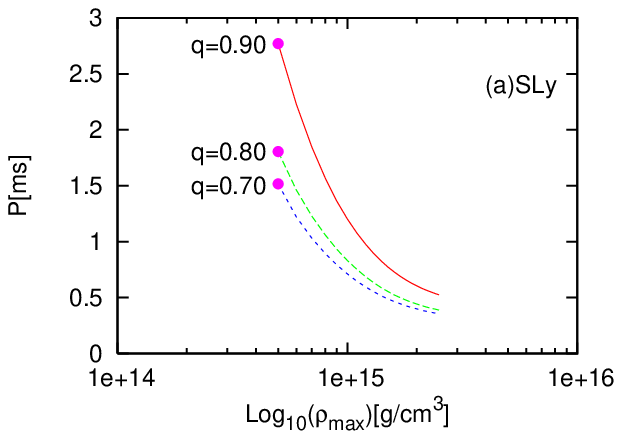}} &
      \resizebox{80mm}{!}{\includegraphics{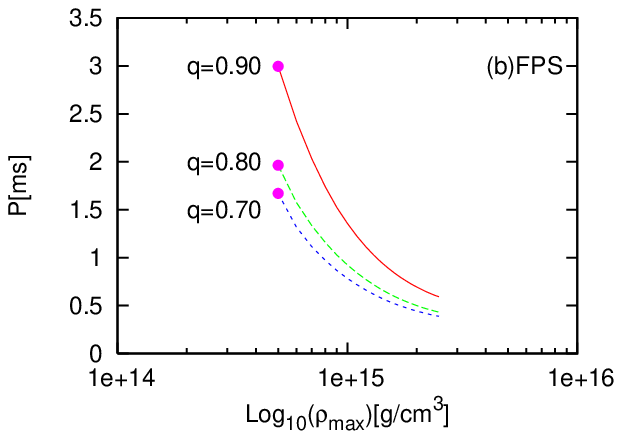}} \\
      \resizebox{80mm}{!}{\includegraphics{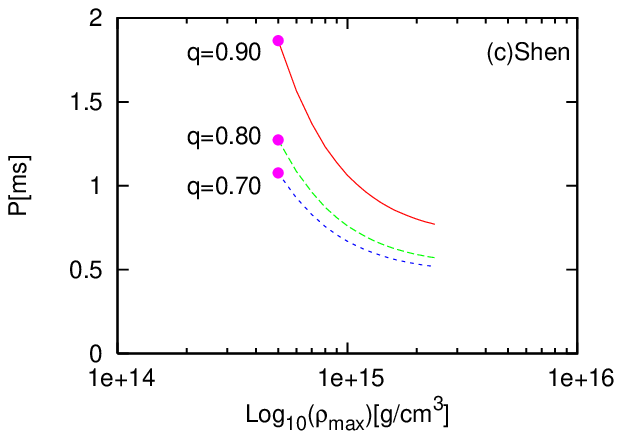}} &
      \resizebox{80mm}{!}{\includegraphics{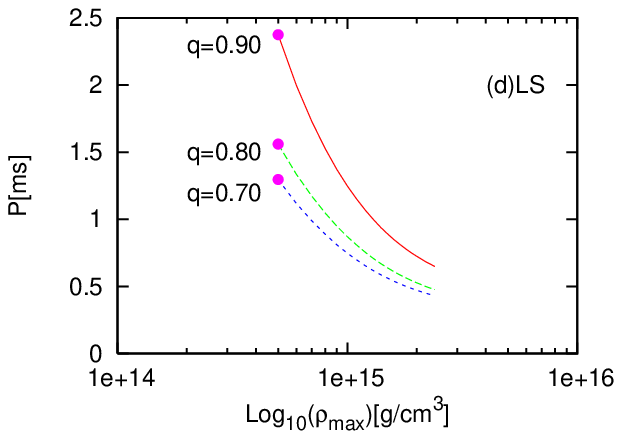}} \\
    \end{tabular}
    \caption{ Rotation periods as functions of the maximum densities 
in the same sequences shown in Figure~\ref{rho-mass}.\label{rho-P}}
  \end{center}
\end{figure*}

\begin{figure*}
  \begin{center}
  \vspace*{174pt}
    \begin{tabular}{cc}
      \resizebox{80mm}{!}{\includegraphics{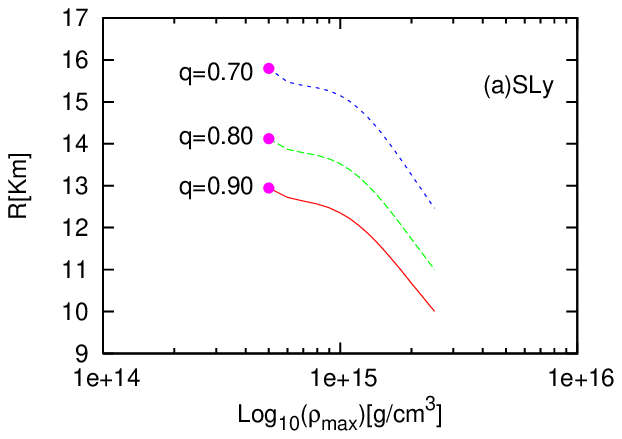}} &
      \resizebox{80mm}{!}{\includegraphics{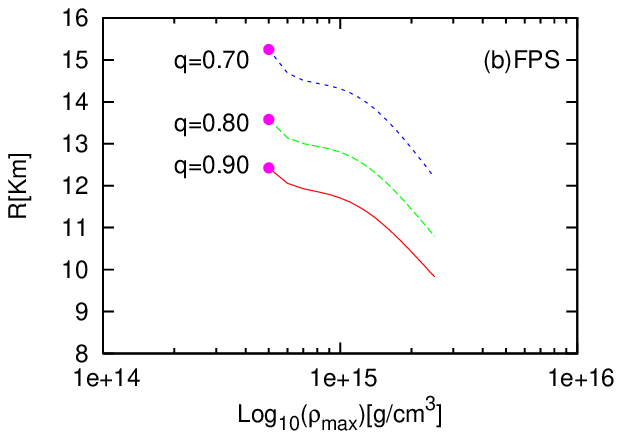}} \\
      \resizebox{80mm}{!}{\includegraphics{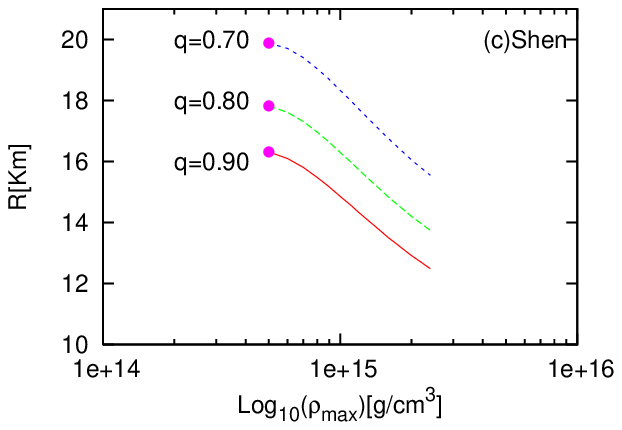}} &
      \resizebox{80mm}{!}{\includegraphics{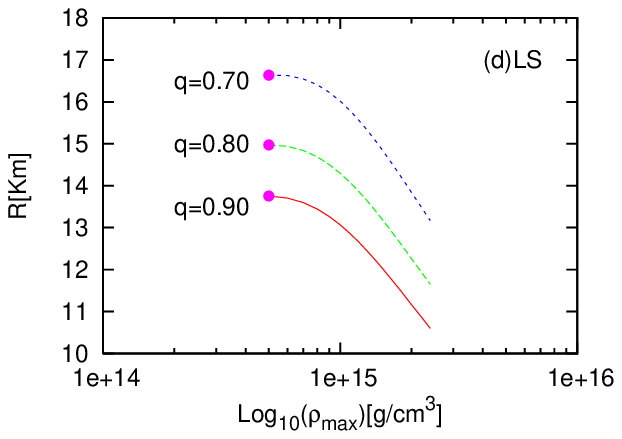}} \\
    \end{tabular}
    \caption{ Equatorial radii as functions of the maximum densities 
in the same sequences shown in Figure~\ref{rho-mass}.\label{rho-R}}
  \end{center}
\end{figure*}

\begin{figure*}
  \begin{center}
  \vspace*{174pt}
    \begin{tabular}{cc}
      \resizebox{80mm}{!}{\includegraphics{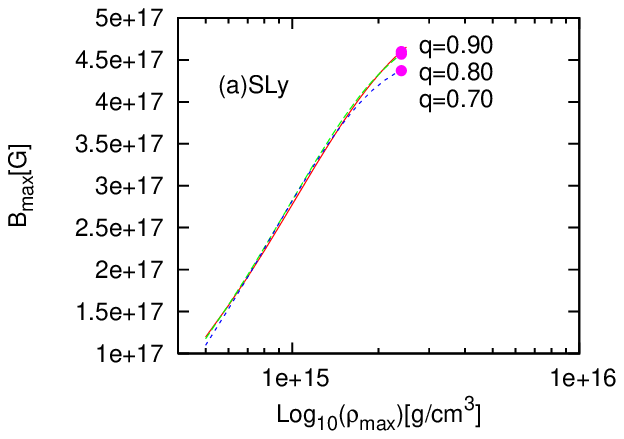}} &
      \resizebox{80mm}{!}{\includegraphics{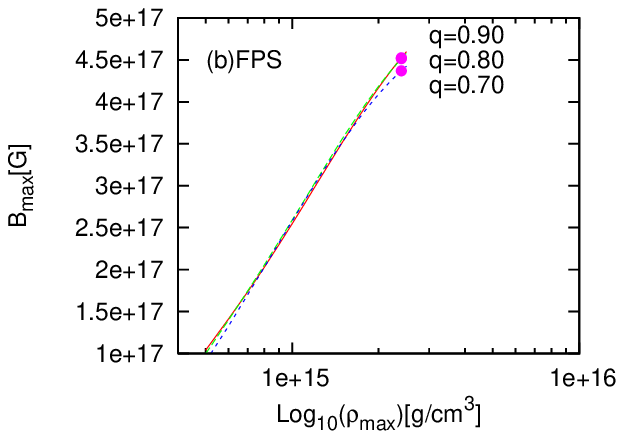}} \\
      \resizebox{80mm}{!}{\includegraphics{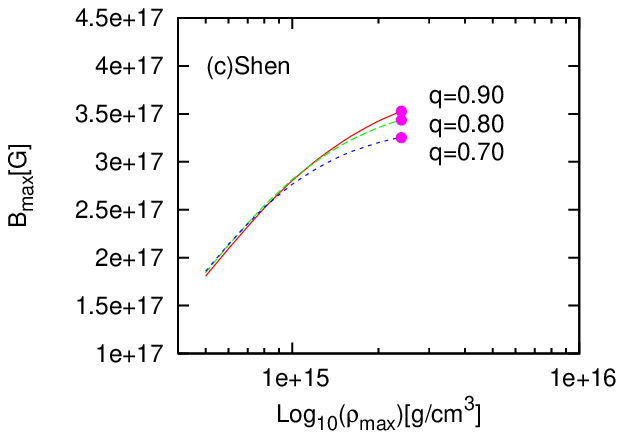}} &
      \resizebox{80mm}{!}{\includegraphics{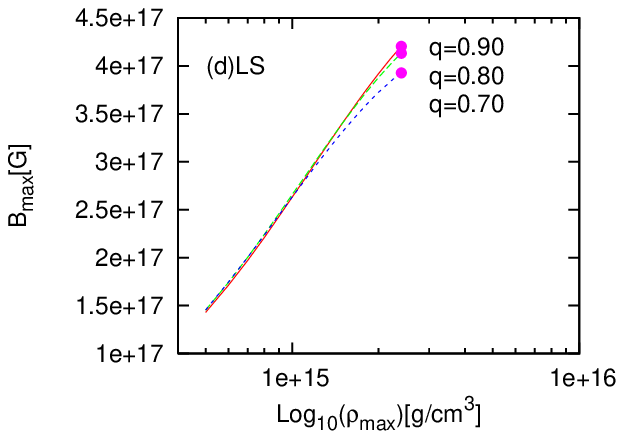}} \\
    \end{tabular}
    \caption{ Maximum magnetic field strength as functions of the 
maximum densities 
in the same sequences shown in Figure~\ref{rho-mass}.\label{rho-Bmax}}
  \end{center}
\end{figure*}

How about relations between other physical quantities and the 
maximum mass ? Figures~\ref{rho-P}, \ref{rho-R}, and \ref{rho-Bmax} are 
the relations between the rotation periods, equatorial radii, and maximum 
magnetic field strength with the maximum density. Figure~\ref{rho-P} shows 
the rotation periods are order of millisecond and they decrease with 
decreasing a value of $q$. This is because these sequences are 
rotation-dominated type (see Table~\ref{gr1}) and more centrifugal force is 
needed to deform the equilibrium star largely. 
These features do not depend on the EOSs. From Figure~\ref{rho-R}, 
the equatorial radii are order of ten kilometer and decreasing functions 
of the maximum density. We also find equatorial radii increase with 
decreasing of $q$. The maximum magnetic field strength are roughly 
$10^{17}[\rm{G}]$ and are increasing functions of maximum density in 
Figure~\ref{rho-Bmax}. Note that 
their values do not depend on $q$ because these sequences belong to 
rotation-dominated type as mentioned above. 

\begin{figure*}
  \begin{center}
  \vspace*{174pt}
    \begin{tabular}{cc}
      \resizebox{80mm}{!}{\includegraphics{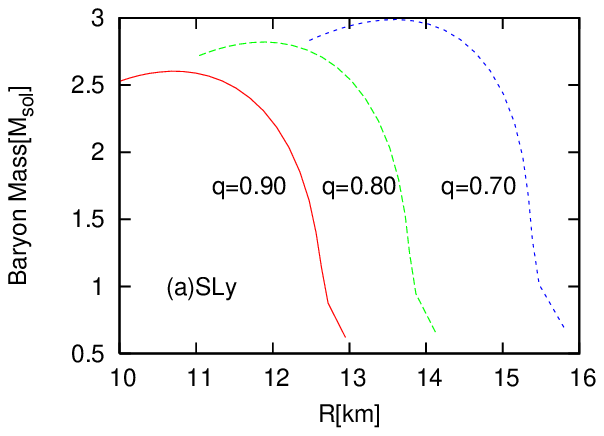}} &
      \resizebox{80mm}{!}{\includegraphics{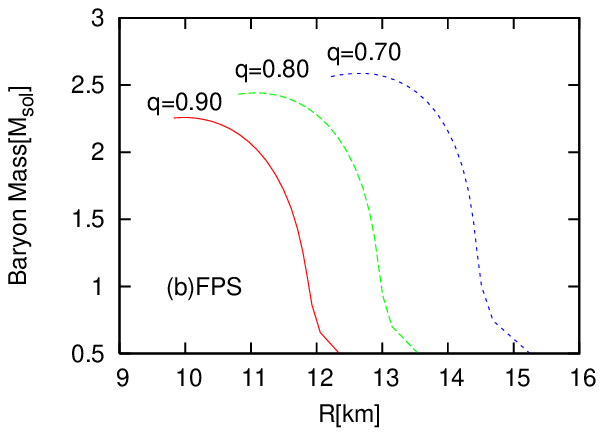}} \\
      \resizebox{80mm}{!}{\includegraphics{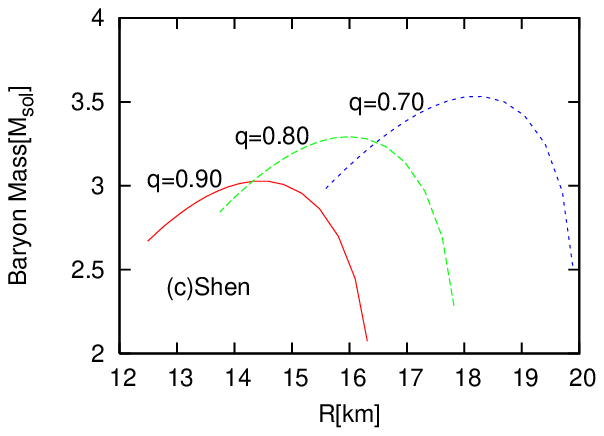}} &
      \resizebox{80mm}{!}{\includegraphics{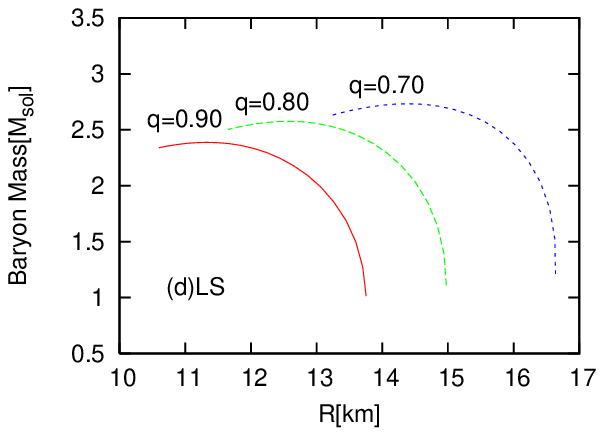}} \\
    \end{tabular}
    \caption{ Relation between baryon mass and equatorial radius 
in the same sequences shown in Figure~\ref{rho-mass}.\label{M-R}}
  \end{center}
\end{figure*}

As we have already shown, the magnetic field near the rotational 
axis is like a dipole decaying as $B_{max}/r^3$ for $r \to \infty$. 
Now, since we find that $B_{max}$ is almost insensitive to the 
values of $q$ with fixing the maximum density, 
$B_p$ should then be proportional to $1/R_p^3$, where $R_p$ is a polar
radius. For instance in the sequences with SLy EOS, $R_p$ changes only
from $11.2[\rm{km}]$ to $10.6[\rm{km}]$ with decreasing of $q$ from $0.9$ 
to $0.7$. On the other hand, it is found that the change in the
equatorial radius is much larger up to about
twenty percents for every EOS employed here when changing $q$ from $0.9$ 
to $0.7$ (see Figure \ref{rho-R}). 
For completeness, we finally show 
$M-R$ relation in Figure~\ref{M-R}, where $M$ and $R$ are baryon mass and 
equatorial radius, respectively. 
It is noted that the relation is also valid for the spherical stars, 
implying that a hierarchy of the stiffness becomes so~\citep{shapiro}.

\subsection{Model of Proto-Magnetar}\label{sec:proto-mag}
In the previous section, we find that the obtained equilibrium configurations
 are difficult to be applied for the observed magnetar.
Here we decide to construct an equilibrium configuration of a
 proto-magnetar bearing in mind a hypothesis by 
 \cite{Duncan:1992} that the magnetars could be originated from the 
proto-neutron-stars with a surface
 magnetic field of order $10^{16}[\rm{G}]$ and with its
 rotation period of an order of milliseconds.
The models computed here should serve as examples of initial models of 
the hydrodynamic/evolutionary simulations of the proto-magnetar.
%

As mentioned in Section~\ref{EOS}, we can construct a model of 
proto-magnetar with finite temperature by employing Shen and LS EOSs.
In the actual calculation, the maximum density, the parameters 
associated with magnetic field, axis ratio and entropy per baryon (in
unit of Boltzmann's constant) are fixed 
as $(\rho_{max},a,\hat{\mu},q,s)=(10^{15}\rm{g}~\rm{cm}^{-3},20,0.1,0.9,2)$.
Here we take the value of the entropy per baryon in the proto-neutron
according to \cite{Prakash:1996xs} indicating its distributions to be
nearly uniform as a result of the convection.

\begin{figure*}
  \begin{center}
  \vspace*{174pt}
    \begin{tabular}{cc}
      \resizebox{80mm}{!}{\includegraphics{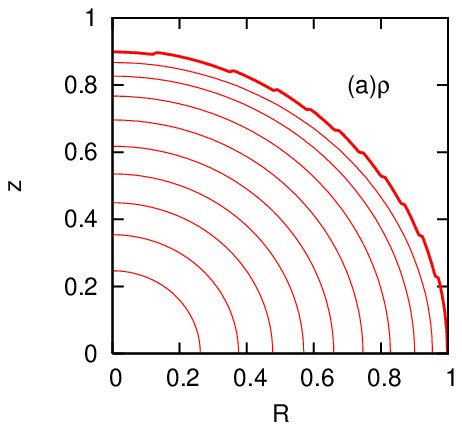}} &
      \resizebox{80mm}{!}{\includegraphics{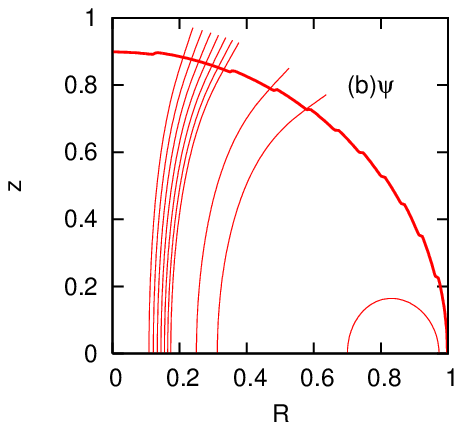}} \\
      \resizebox{80mm}{!}{\includegraphics{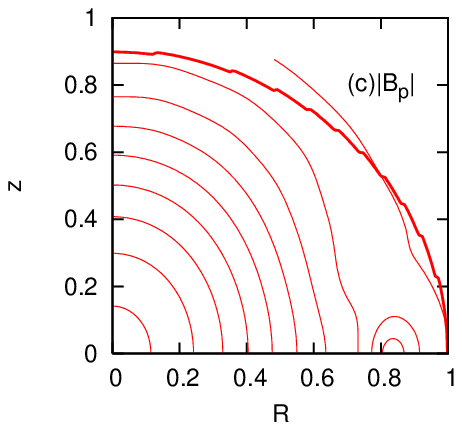}} &  
      \resizebox{80mm}{!}{\includegraphics{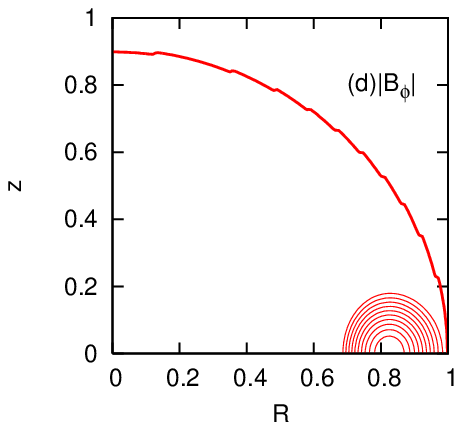}} \\
    \end{tabular}
    \caption{(a) Density distribution, (b) magnetic field line (c) poloidal 
magnetic field (d) toroidal magnetic field in proto-magnetar equilibrium 
configuration with Shen EOS.\label{Shen-protomag}}
  \end{center}
\end{figure*}

\begin{figure*}
  \begin{center}
  \vspace*{174pt}
    \begin{tabular}{cc}
      \resizebox{80mm}{!}{\includegraphics{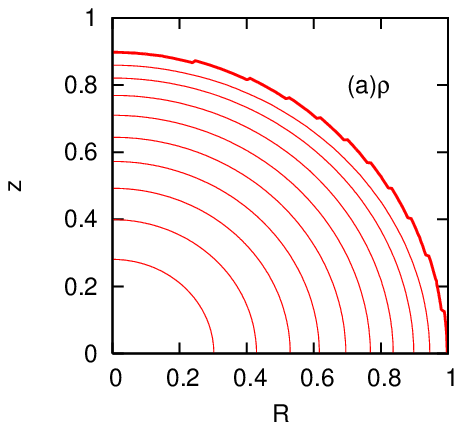}} &
      \resizebox{80mm}{!}{\includegraphics{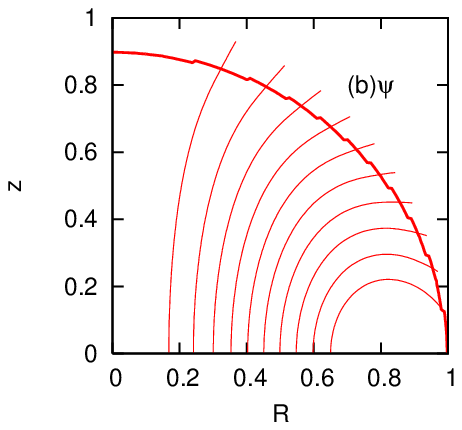}} \\
      \resizebox{80mm}{!}{\includegraphics{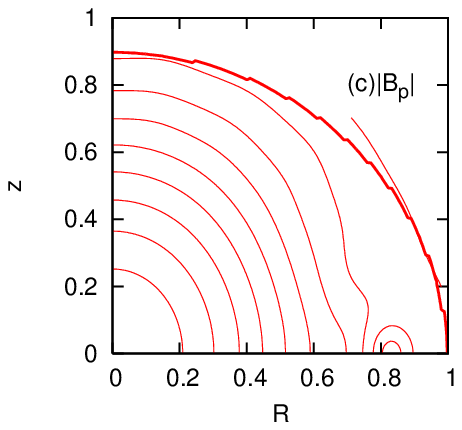}} &  
      \resizebox{80mm}{!}{\includegraphics{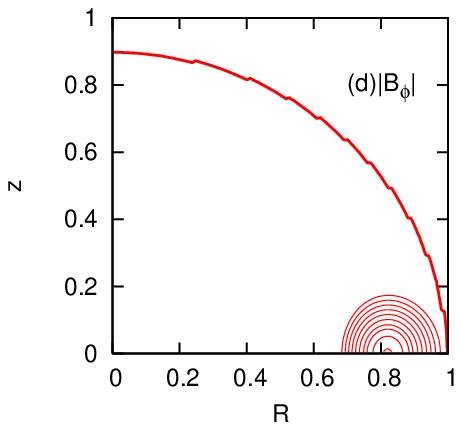}} \\
    \end{tabular}
    \caption{(a) Density distribution, (b) magnetic field line (c) poloidal 
magnetic field (d) toroidal magnetic field in proto-magnetar equilibrium 
configuration with LS EOS.\label{LS-protomag}}
  \end{center}
\end{figure*}

Results are summarized in Table~\ref{hotmag}. Note that magnetic field 
strength at pole is order of $10^{16}[\rm{G}]$. Matter and magnetic field 
structure of these stars are depicted in Figure~\ref{Shen-protomag} 
and \ref{LS-protomag}. As a result, it is found that 
 the features appearing in the cold magnetized equilibrium 
configurations above are still valid despite of the incursion of the finite
temperature effect on the EOSs.
More interestingly, the masses of magnetars are also dependent
on the EOSs as shown in
Table~\ref{hotmag}. They are predicted to be as large as $3.0
M_{\odot}$ for the LS EOS, such neutron stars have never been observed 
\citep{Lattimer:2006}.
Although we have little observational
information about the masses of magnetar so far, our results suggest
that we may obtain information about the EOS from the observation of
the masses of magnetars.

\begin{table*}
\centering
 \begin{minipage}{140mm}
\caption{Equilibrium configuration of hot neutron star with Shen and 
LS EOS for $(\rho_{max},a,\hat{\mu},q,s)
=(10^{15}\rm{g}~\rm{cm}^{-3},20,0.1,0.9,2)$, 
where $s$ is the entropy per baryon in unit of Boltzmann's constant.\label{hotmag}}
\begin{tabular}{ccccccc}
\hline\hline
$q$ & $Bp[\rm{G}]$ & $Bc[\rm{G}]$ & $M[M_\odot]$ & $P[\rm{ms}]$ 
& $R[\rm{km}]$ & h \\
\hline
\multicolumn{7}{c}{Shen}\\
\hline
0.90&6.73E+16&2.63E+17&1.99E+01&1.25E+00&13.1E+00&0.367E+00\\
\hline
\multicolumn{7}{c}{LS}\\
\hline
0.90&7.32E+16&2.80E+17&3.01E+01&1.06E+00&14.9E+00&0.326E+00\\
\hline
\end{tabular}
\end{minipage}
\end{table*}
\clearpage

\clearpage
\section{Conclusion}\label{conc}
In this paper, we have investigated equilibrium sequences of magnetized rotating stars
with four kinds of realistic equations of state (EOSs) of SLy \citep{douchin}
et al.), FPS \citep{pand}, Shen \citep{shen98}, and LS \citep{LS}.
We have employed the Tomimura-Eriguchi
scheme to construct the equilibrium configurations.

At first, we have obtained the solution in the regime of 
Newtonian gravity. Basic properties of the magnetized 
Newtonian stars are summarized as follows: 
(1) For the non-rotating sequences, 
there exist nearly toroidal configurations, $q_{min} \sim 0$, irrespective of 
the EOSs. The magnetic energy stored in the stars increases with the
degree of deformation being larger.
(2) 
For the rotating sequences, we have categorized the sequences with 
four kinds of EOSs as rotation-dominated (RD) type, magnetic-dominated 
(MD) one, and 
nearly toroidal one. As a result, the sequences
with softer EOSs of SLy 
and FPS are found to belong to RD or MD type and the sequences with
stiffer EOSs of Shen and LS can belong also to the nearly toroidal one.
 
(3) We have also focused on the structure of equilibrium configurations.
Reflecting the stiffness of EOSs,
 the density distributions of stars with SLy or FPS EOSs concentrate
more at the center than those of the stars with Shen or 
LS EOSs. The toroidal fields
for SLy and FPS EOSs are found to distribute in relatively wider regions in the
vicinity of the equatorial plane than for
 Shen and LS EOSs. The poloidal magnetic 
fields are also affected by the toroidal fields because poloidal fields are 
highly distorted where the toroidal fields are strong.
Regardless of the difference of the EOSs, 
a global configuration of magnetic field line is found to be universal, namely the tori of twisted field lines 
around the symmetry axis inside the star and the untwisted poloidal 
field lines, which penetrate the stellar surface to continue to the
vacuum. 

Then, adding the GR correction to the gravity, we have performed the quantitative 
investigation of the strong magnetized equilibrium stars. 
 As a result, we find that the
difference due to the EOSs becomes small because all the employed EOSs
become sufficiently stiff for the large maximum density, typically
greater than $10^{15}\rm{g}~\rm{cm}^{-3}$.
We have investigated the relation between the baryon mass, 
the magnetic field at pole, 
the rotational period, the equatorial radius, and the maximum magnetic 
field as a function of the maximum density. 
The typical magnetic fields at pole 
are about $10^{16} \rm{G}$, the periods are about several milliseconds, 
the radii are about ten kilometers
and the maximum magnetic fields are about $10^{17} \rm{G}$.
The maximum mass is found
to be $3.0 M_\odot$ for SLy 
EOS, $2.6 M_\odot$ for FPS EOS, $3.5 M_\odot$ for Shen EOS
and $2.7 M_\odot$ for LS EOS for $q=0.7$ configurations, respectively. 
These values are about twenty percents increasing for that of the 
spherical stars. 

Finally, we have computed equilibrium sequences at finite
temperature for the Shen and LS EOSs aiming to construct the
equilibrium configurations of the proto-magnetars.
As a result, it is found that 
 the features appearing in the cold magnetized equilibrium 
configurations above are still valid despite of the incursion of the finite
temperature effect on the EOSs.
 Since the masses of the proto-magnetars are highly
dependent on the EOSs, we have speculated that one may obtain
information about the EOSs from the observation of
the masses of magnetars.

It is true that our treatment for the general relativistic effect is 
nothing but a crude approximation. Thus we consider this study as a prelude
to a fully general relativistic study, however hoping for the moment that 
our results could serve as the initial condition for the 
hydrodynamic evolutionary studies of newly-born magnetars including
the microphysical EOSs.

\section*{Acknowledgments}
We express thanks to S. Yoshida for fruitful discussions and to 
H. Ono, H. Suzuki, and K. Sumiyoshi for providing us the tabulated
table for Lattimer-Swesty EOS.
Kiuchi thanks to K.~i. Maeda and Kotake to K. Sato for continuing encouragements.
Kotake also thanks to S. Yamada and E. M\"{u}ller for informative discussions. 
 This work was supported in part by the Japan Society for 
Promotion of Science(JSPS) Research Fellowships (Kiuchi), 
Grants-in-Aid for the Scientific Research from the Ministry of
Education, Science and Culture of Japan (No.1840044 and S19104006).

\clearpage
\appendix

\section{Code Test}\label{code test}
To check our code ability, we calculate the same sequences shown in 
\cite{yos} with polytropic equations of state. 
We actually obtain two kinds of 
sequences, non-rotating sequence and rotating-sequence. 
In Table~\ref{non-rot}, some physical quantities of the non-rotating 
sequence with the polytropic index $N=3.0$, $k=0.1$ and 
$a=15$ (see equation~(\ref{eq26}), (\ref{eq27})) are shown. 
For the axis ratio $q$, the upper row corresponds to the result by 
\cite{yos} and the down one do to our result.
Table~\ref{rig-rot} is a result of the rotating sequence with polytropic 
index $N=0.5$, $a=20$, and $\hat{\mu}=0.05$. 
In all sequences, the relative 
errors for the each quantities 
with \cite{yos}'s result are less than one percent and VC is 
$10^{-4}\sim10^{-5}$. Therefore, we confirm that our code works well.
\begin{table*}
\centering
\begin{minipage}{140mm}
\caption{Physical quantities for the non-rotating sequence with 
polytropic index $N=3$, $a=15$, $k=0.1$,
 and $\hat{\Omega}^2=0$. For the axis ratio, the upper row represents 
Yoshida \& Eriguchi's result~\citep{yos}
 and the down one does our result. \label{non-rot}}
\begin{tabular}{ccccccccc}
\hline\hline
q & $H/|W|$ & $U/|W|$ & $|\hat{W}|$ & $\hat{\mu}$ & $\hat{C}$ & 
$\hat{\beta}$ & $\hat{M}$ & VC \\
\hline
0.98 &0.0050 &0.332 &7.310E-4 &0.140 &-6.348E-3 &5.276E-3 &0.078 &\\
     &0.0050 &0.332 &7.310E-4 &0.140 &-6.348E-3 &5.276E-3 &0.078 &2.55E-4\\
0.90 &0.0275 &0.324 &7.922E-4 &0.299 &-7.163E-3 &5.303E-2 &0.082 &\\
     &0.0275 &0.324 &7.922E-4 &0.299 &-7.163E-3 &5.303E-2 &0.082 &2.54E-4\\
0.80 &0.0635 &0.312 &9.004E-4 &0.408 &-8.526E-3 &5.335E-2 &0.088 &\\
     &0.0635 &0.312 &9.004E-4 &0.408 &-8.526E-3 &5.335E-2 &0.088 &2.50E-4\\
0.70 &0.1131 &0.296 &1.097E-3 &0.476 &-1.064E-2 &5.400E-3 &0.099 &\\
     &0.1131 &0.296 &1.097E-3 &0.476 &-1.064E-2 &5.400E-3 &0.099 &2.37E-4\\
0.60 &0.1856 &0.272 &1.560E-3 &0.501 &-1.456E-2 &5.576E-3 &0.121 &\\
     &0.1856 &0.272 &1.560E-3 &0.501 &-1.456E-2 &5.575E-3 &0.121 &2.07E-4\\ 
0.50 &0.3044 &0.232 &4.087E-3 &0.447 &-2.721E-2 &6.524E-3 &0.211 &\\
     &0.3042 &0.232 &4.079E-3 &0.447 &-2.718E-2 &6.520E-3 &0.210 &1.21E-4\\
0.40 &0.4087 &0.197 &7.624E-3 &0.279 &-3.845E-2 &6.188E-3 &0.320 &\\
     &0.4086 &0.197 &7.637E-3 &0.279 &-3.849E-2 &6.190E-3 &0.321 &6.59E-6\\
0.30 &0.4219 &0.193 &4.626E-3 &0.247 &-3.022E-2 &4.713E-3 &0.254 &\\
     &0.4218 &0.193 &4.638E-3 &0.247 &-3.026E-2 &4.716E-3 &0.254 &8.43E-6\\
0.20 &0.4287 &0.190 &3.434E-3 &0.237 &-2.635E-2 &4.003E-3 &0.220 &\\
     &0.4287 &0.190 &3.453E-3 &0.237 &-2.643E-2 &4.009E-3 &0.221 &3.12E-6\\
0.10 &0.4326 &0.189 &2.900E-3 &0.233 &-2.443E-2 &3.648E-3 &0.203 &\\
     &0.4326 &0.189 &2.926E-3 &0.233 &-2.454E-2 &3.658E-3 &0.204 &1.52E-5\\
0.01 &0.4338 &0.189 &2.743E-3 &0.233 &-2.384E-2 &3.539E-3 &0.197 &\\
     &0.4338 &0.189 &2.771E-3 &0.232 &-2.396E-2 &3.550E-3 &0.198 &1.96E-5\\
\hline
\end{tabular}
\end{minipage}
\end{table*}
\begin{table*}
\centering
 \begin{minipage}{140mm}
\caption{Same as Table~\ref{non-rot}, but for the rotating sequence 
with polytropic index $N = 0.5$, $a=20$, $k=0,1$, and $\hat{\mu}=0.05$. 
\label{rig-rot}}
\begin{tabular}{cccccccccc}
\hline\hline
q & $H/|W|$ & $U/|W|$ & $T/|W|$ & $|\hat{W}|$ &$\hat{\Omega}^2$ & $\hat{C}$ & 
$\hat{\beta}$ & $\hat{M}$ & VC \\
\hline
$\hat{\Omega}^2<0$&-&-&-&-&-&-&-&-&-\\
0.98&2.75E-3&0.331&2.04E-3&2.68E-1&1.50E-3&-0.181&8.61E-2&2.24&\\
    &2.74E-3&0.331&2.04E-3&2.68E-1&1.50E-3&-0.181&8.61E-2&2.24&2.19E-4\\
0.90&2.88E-3&0.317&2.37E-2&2.31E-1&1.64E-2&-0.175&7.78E-2&2.05&\\
    &2.89E-3&0.317&2.37E-2&2.31E-1&1.64E-2&-0.175&7.77E-2&2.05&1.37E-4\\
0.80&3.10E-3&0.297&5.33E-2&1.86E-1&3.40E-2&-0.166&6.71E-2&1.80&\\
    &3.10E-3&0.297&5.33E-2&1.86E-1&3.40E-2&-0.166&6.71E-2&1.80&1.25E-4\\
0.70&3.37E-3&0.275&8.60E-2&1.44E-1&4.98E-2&-0.155&5.63E-2&1.55&\\
    &3.37E-3&0.274&8.60E-2&1.44E-1&4.98E-2&-0.155&5.63E-2&1.55&1.35E-4\\
0.60&3.72E-3&0.251&1.22E-1&1.04E-1&6.29E-2&-0.141&4.55E-2&1.29&\\
    &3.73E-3&0.251&1.22E-1&1.04E-1&6.29E-2&-0.141&4.55E-2&1.29&1.22E-4\\
0.50&4.21E-3&0.225&1.61E-1&6.75E-2&7.19E-2&-0.123&3.46E-2&1.01&\\
    &4.22E-3&0.225&1.61E-1&6.76E-2&7.19E-2&-0.123&3.46E-2&1.01&8.79E-5\\
0.43&3.64E-3&0.210&1.83E-1&4.06E-2&7.46E-2&-0.101&2.67E-2&0.750&\\
    &3.67E-3&0.210&1.84E-1&4.07E-2&7.46E-2&-0.102&2.67E-2&0.750&9.53E-5\\
MS&-&-&-&-&-&-&-&-&-\\
\hline
\end{tabular}
\end{minipage}
\end{table*}

\section{General Relativistic Correction}\label{GR correct}
\begin{figure*}
  \begin{center}
  \vspace*{40pt}
    \begin{tabular}{cc}
      \resizebox{80mm}{!}{\includegraphics{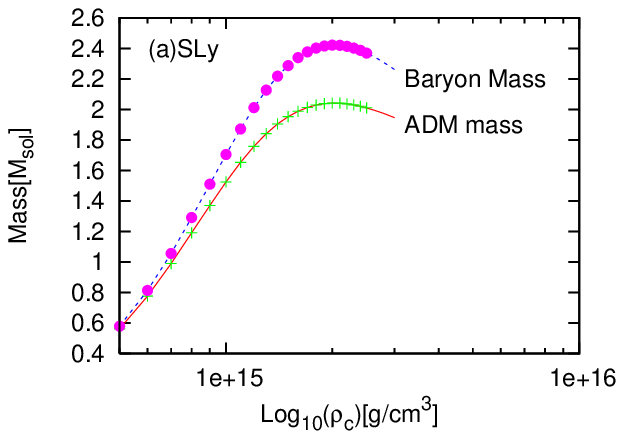}} &
      \resizebox{80mm}{!}{\includegraphics{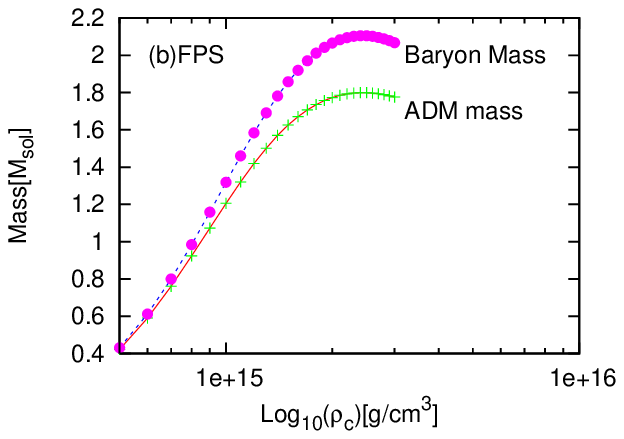}} \\
      \resizebox{80mm}{!}{\includegraphics{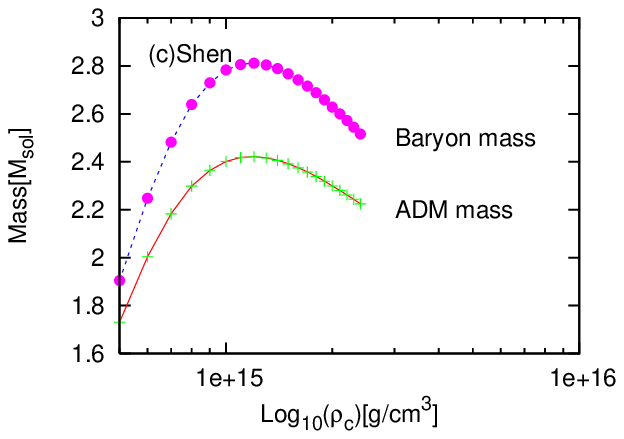}} &
      \resizebox{80mm}{!}{\includegraphics{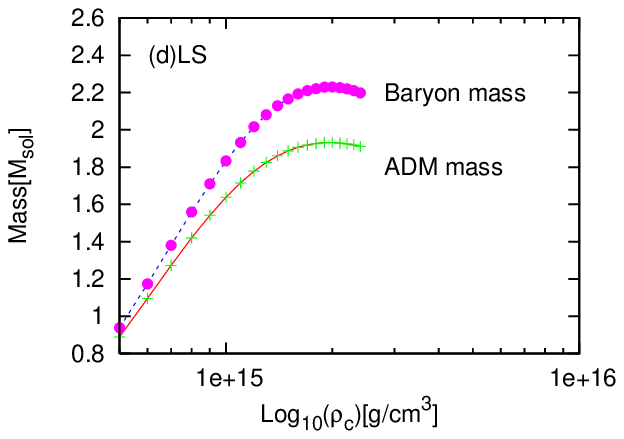}}  \\
    \end{tabular}
    \caption{ADM mass and baryon mass of spherically symmetric stars with 
SLy, FPS, Shen, or LS EOSs as a function of central density.
The solid and dotted lines are the results of TOV code and 
the cross symbols and filled circles are those of 2D code, respectively.
\label{GR1D}}
  \end{center}
\end{figure*}

We start from the metric of spherical symmetry space time,
\begin{eqnarray}
ds^2 = - c^2e^{2\Phi_{GR}}dt^2 + e^{2\Lambda}dr^2 
+ r^2 ( d\theta^2 + \sin^2 \theta d\varphi^2 ).\label{metric}
\end{eqnarray}
Matter is assumed as a perfect fluid, thus the energy momentum tensor is
\begin{eqnarray}
T^{\mu\nu} = \left( \rho ( 1 + e ) + \frac{P}{c^2} 
\right) u^\mu u^\nu + P g^{\mu\nu},
\label{pfluid}
\end{eqnarray}
where $\rho$, $e$, and $P$ are a density, specific internal energy, 
and pressure. TOV equation~\citep{opp} is
\begin{eqnarray}
&&\frac{dm}{dr} = 4 \pi r^2 \rho ( 1 + e ),\label{TOV1}\\
&&\frac{dP}{dr} = - G \frac{\rho h}{r^2}( m + \frac{4\pi r^3 P}{c^2} )
/( 1 - \frac{2Gm}{c^2r} ),\label{TOV2},
\end{eqnarray}
where $h=1+e+P/\rho c^2$ is relativistic enthalpy and $m$ is mass function 
defined as
\begin{eqnarray*}
e^{2\Lambda} = \frac{1}{1-\frac{2Gm}{c^2r}}.
\end{eqnarray*}
Equation for the potential $\Phi_{GR}$ is
\begin{eqnarray}
\frac{d\Phi_{GR}}{dr} = \frac{G}{c^2r^2}\left( m + \frac{4\pi r^3 P}{c^2} 
\right)/( 1 - \frac{2Gm}{c^2r} ).\label{TOV3}
\end{eqnarray}
This potential, of course, reduces to the gravitational potential 
$\Phi$ in the Newtonian limit. 
To make clear a general relativistic contribution 
in the potential, we perform the Taylor expansion of equation~(\ref{TOV3}) 
with respect to $x(r) \equiv 2Gm(r)/r c^2$, 
which is guaranteed 
to be small than unity in star. Result is written in the form,
\begin{eqnarray}
&&\frac{d\Phi_{GR}}{dr} = \frac{G}{c^2 r^2} \int^r_0 4\pi r^2 \rho dr' 
\nonumber\\
&&+ \frac{1}{r} \Big[ \frac{G}{c^2 r}\int^r_0 4 \pi r^2 \rho e dr'
\nonumber\\
&&+ \frac{1}{2}\sum^\infty_{n=2} x(r)^n 
  + \frac{4\pi G r^2 P}{c^4} \sum^\infty_{n=0} x(r)^n \Big]. \label{TOV4}
\end{eqnarray}
Note that the first term is a contribution from Newtonian gravitational 
potential. So, we define a general relativistic correction in the potential 
as $\delta \Phi_{GR} \equiv c^2 \Phi_{GR} - \Phi_N $. 
Boundary condition is derived from a requirement that the outer vacuum space 
time is equivalent to the Schwarzschild space time,
\begin{eqnarray}
\Phi_{GR} = \frac{1}{2} \ln \left( 1 - \frac{2GM}{c^2 R} \right),\label{TOV5a}
\end{eqnarray}
where $R$ and $M$ are radius and ADM mass of the star. From the 
Taylor expansion 
with respect to $x(R)=\frac{2GM}{c^2R}$, the boundary condition for the 
general relativistic correction term $\delta \Phi_{GR}$ is derived. 
As a result, $\delta \Phi_{GR} $ is expressed in an integral form, 
\begin{eqnarray}
&&\delta \Phi_{GR}(r) = - \frac{c^2}{2} \sum^\infty_{n=2} \frac{x(R)^n}{n} - 
\frac{G}{r} \int^r_0 4 \pi \tilde{r}^2 \rho e d\tilde{r} \nonumber\\
&&- \int^R_r d\tilde{r} \frac{1}{\tilde{r}}\Big[
4 \pi G \tilde{r}^2 \rho e + \frac{c^2}{2} \sum^\infty_{n=2} x(\tilde{r})^n \nonumber\\
&&+ \frac{4\pi G r^2 P}{c^2} \sum^\infty_{n=0} x(\tilde{r})^n \Big].\label{TOV5b}
\end{eqnarray}
This spherically symmetric effective potential is also applied in our 2D 
axisymmetric configuration code, where we first compute angular averages 
of the relevant hydrodynamical variables. These are then used to calculate 
the spherical general correction term $\delta \Phi^{1D}_{GR}(r)$ in 
equation~(\ref{TOV5b}). Consequently, we modify the 2D Newtonian potential 
$\Phi^{2D}_N(r,\theta)$ to obtain the two dimensional general relativistic 
potential
\begin{eqnarray}
\Phi^{2D}_{GR}(r,\theta) = \Phi^{2D}_{N}(r,\theta) + \delta \Phi^{1D}_{GR}(r).
\label{TOV5c}
\end{eqnarray}
The method shown in \cite{mar} has an ambiguity in definition of potential 
and imposes boundary condition of their potential at infinity. However, 
in our method the general relativistic correction in the potential is 
explicitly written down and its boundary condition is imposed at the stellar 
surface. So, the two method is a little bit different. Verification of our 
method is shown in the last in this appendix. 
Subsequently, we focus on the Bernoulli equation, which 
is ordinary used to obtain axisymmetric configuration (e.g.~(\ref{eq22})). 
Under the metric form~(\ref{TOV1}), four velocity is given by 
\begin{eqnarray}
u^\mu = ( e^{-\Phi_{GR}},0,0,0).\label{TOV6}
\end{eqnarray}
The relativistic Bernoulli equation with spherically symmetric is also 
written as
\begin{eqnarray}
c^2 \ln h = \ln u^t + C,\label{TOV7}
\end{eqnarray}
where $h$ and C are relativistic enthalpy, integral constant, respectively. 
Combining equations~(\ref{TOV6})-(\ref{TOV7}), we find 
\begin{eqnarray}
c^2 \ln h = - (\Phi_N + \delta \Phi_{GR}) + C.\label{TOV8}
\end{eqnarray}
Finally we modify the Bernoulli equation of magnetized rotating case as 
\begin{eqnarray}
c^2 \ln h = - \Phi^{2D}_{GR} + \frac{1}{2} R^2 \Omega^2 + 
\int^{\Psi} \mu(u) du + C. \label{TOV9}
\end{eqnarray}
Note that this treatment may be speculation.
To verify our treatment, we calculate spherical symmetric 
stars with four kinds of EOS in both 1D TOV code and 2D axisymmetric code. 
Relations of mass, both ADM and baryon mass, with central density 
in these stars are displayed in Figure~\ref{GR1D}. 
We confirm our treatment works well.

\end{document}